\documentclass[twocolumn]{aastex63}

\graphicspath{{./}{figures/}}

\usepackage{booktabs}
\usepackage{comment}
\usepackage{fourier}

\newcommand{\overbar}[1]{\mkern 1.5mu\overline{\mkern-1.5mu#1\mkern-1.5mu}\mkern 1.5mu}

\shorttitle{Cosmic SZ Background}
\shortauthors{Chiang et al.}

\begin{document}

\title{The Cosmic Thermal History Probed by Sunyaev-Zeldovich Effect Tomography}

\author[0000-0001-6320-261X]{Yi-Kuan Chiang}
\affiliation{Center for Cosmology and AstroParticle Physics (CCAPP), The Ohio State University, Columbus, OH 43210, USA}

\author[0000-0001-5133-3655]{Ryu Makiya}
\affiliation{Kavli Institute for the Physics and Mathematics of the Universe (Kavli IPMU, WPI), University of Tokyo, Chiba 277-8582, Japan}

\author[0000-0003-3164-6974]{Brice M\'enard}
\affiliation{Department of Physics \& Astronomy, Johns Hopkins University, 3400 N. Charles Street, Baltimore, MD 21218, USA}
\affiliation{Kavli Institute for the Physics and Mathematics of the Universe (Kavli IPMU, WPI), University of Tokyo, Chiba 277-8582, Japan}

\author[0000-0002-0136-2404]{Eiichiro Komatsu}
\affiliation{Max-Planck-Institut f\"{u}r Astrophysik, Karl-Schwarzschild Str. 1, 85741 Garching, Germany}
\affiliation{Kavli Institute for the Physics and Mathematics of the Universe (Kavli IPMU, WPI), University of Tokyo, Chiba 277-8582, Japan}

\begin{abstract}
The cosmic thermal history, quantified by the evolution of the mean thermal energy density in the universe, is driven by the growth of structures as baryons get shock heated in collapsing dark matter halos. This process can be probed by redshift-dependent amplitudes of the thermal Sunyaev-Zeldovich (SZ) effect background. To do so, we cross-correlate eight sky intensity maps in the $\it{Planck}$ and Infrared Astronomical Satellite missions with two million spectroscopic redshift references in the Sloan Digital Sky Surveys. This delivers snapshot spectra for the far-infrared to microwave background light as a function of redshift up to $z\sim3$. We decompose them into the SZ and thermal dust components. Our SZ measurements directly constrain $\langle bP_{\rm e} \rangle$, the halo bias-weighted mean electron pressure, up to $z\sim 1$. This is the highest redshift achieved to date, with uncorrelated redshift bins thanks to the spectroscopic references. We detect a threefold increase in the density-weighted mean electron temperature $\overbar{T}_{\rm{e}}$ from $7\times 10^5~{\rm K}$ at $z=1$ to $2\times 10^6~{\rm K}$ today. Over $z=1$--$0$, we witness the build-up of nearly $70\%$ of the present-day mean thermal energy density $\rho_{\rm{th}}$, with the corresponding density parameter $\Omega_{\rm th}$ reaching $1.5 \times10^{-8}$. We find the mass bias parameter of $\it{Planck}$'s universal pressure profile of $B=1.27$ (or $1-b=1/B=0.79$), consistent with the magnitude of non-thermal pressure in gas motion and turbulence from mass assembly. We estimate the redshift-integrated mean Compton parameter $y\sim1.2\times10^{-6}$, which will be tested by future spectral distortion experiments. More than half of which originates from the large-scale structure at $z<1$, which we detect directly.
\end{abstract}

\

\section{Introduction} \label{sec:intro}

One of the major outcomes of cosmological structure formation is the deepening of gravitational potential seeded in primordial density fluctuations, and the subsequent conversion of gravitational energy into thermal energy in collapsed structures \citep{1999ApJ...514....1C,2004ApJ...616..643F}. During this process, an increasing fraction of cosmic baryons are accreted onto dark matter halos and are shock heated to the virial temperature ($10^5$--$10^8$ K). Measurements of the cosmic hot gas content thus directly probe the growth of structure and the thermalization process.

The thermal Sunyaev-Zeldovich (tSZ) effect \citep{SZ_1972} is a powerful probe of hot baryons in the universe. It appears as a spectral distortion of the cosmic microwave background (CMB) as the CMB photons are inverse-Compton scattered off of free electrons in the intervening gas \citep[see][for reviews]{2002ARA&A..40..643C,kitayama:2014,2019SSRv..215...17M}. The amplitude of the tSZ effect, namely the Compton $y$ parameter, scales with the electron pressure integrated along the line of sight. As a result, the global, cosmic tSZ signal is dominated by massive structures: clusters and groups at low redshifts and protoclusters at high redshifts \citep{2000PhRvD..61l3001R,2001PhRvD..63f3001S,2002MNRAS.336.1256K,2012ApJ...758...75B,2013ApJ...779..127C,2015PhRvL.115z1301H,dolag/komatsu/sunyaev:2016,2016A&ARv..24...14O}.

There are several advantages of the tSZ effect over direct X-ray Bremsstrahlung emission in probing the cosmic mean thermal history. First, the amplitude of the tSZ signal scales directly with the thermal energy content of the universe, while that of the X-ray emission scales with gas density squared and less with temperature. The X-ray emission is thus more affected by the clumpiness of gas.  Second, the characteristic spectral dependence of the tSZ signal allows for a robust extraction against other sources of radiation, while X-ray emissions of different origins show roughly featureless power-law or exponential spectra. X-ray line emission helps but demands high photon statistics, which is not available for study of the global extragalactic X-ray background. Finally, the amplitude of the tSZ signal does not suffer from the $(1+z)^{-4}$ surface brightness dimming, potentially allowing us to probe its evolution over a wide range of cosmic time. 

Having an efficiency independent of redshift also comes with a challenge: the spectral features of the tSZ effect do not inform the redshifts of the sources. To probe the growth history of structure with the tSZ effect, external redshift information is needed \citep{2001ApJ...549...18Z, 2011ApJ...730..127S}. One way to deproject the contribution of the tSZ Compton $y$ parameter (or any intensity field on the sky) along the line of sight is the so-called clustering-based redshift inference \citep{Newman_2008, McQuinn_White_2013, Menard_2013}. In this approach, one takes an external sample of reference sources with known redshifts and measures spatial cross-correlations with the intensity field as a function of redshift. Correlated background intensities can then be extracted tomographically \citep{Schmidt_2015, Chiang_Menard_2019, Chiang_2019}. This technique has also been used to estimate redshift distributions of discrete sources \citep[e.g.,][]{2015MNRAS.447.3500R, 2016MNRAS.457.3912R, 2016MNRAS.460..163R}. 

Recently, \cite{2019PhRvD.100f3519P} estimated the mean tSZ history up to $z=0.7$ by cross-correlating {\it Planck}-based $y$ maps \citep{planck2016_sz_map} and galaxies in the Dark Energy Survey \citep{2018PhRvD..98d3526A} with photometric redshifts. \cite{2020MNRAS.491.5464K} performed a similar analysis with photometric redshifts in the 2MASS and WISE$\,\times\,$SuperCOSMOS galaxy catalogs \citep{2014ApJS..210....9B, 2016ApJS..225....5B}. Other works have also explored the cosmological and astrophysical information in tSZ--galaxy cross-correlations or stacking \citep{2004PhRvD..69h3524A,2013A&A...557A..52P, 2014MNRAS.445..460G, 2015ApJ...808..151G, ruan/mcquinn/anderson:2015, crichton/etal:2016, spacek/etal:2016, spacek/etal:2017, soergel/etal:2017, 2017MNRAS.467.2315V, 2018PhRvD..97f3514A, 2018ApJ...854..181L, 2018MNRAS.480.4017L, 2018MNRAS.480.3928M, 2019A&A...624A..48D, hall/etal:2019, 2019MNRAS.483..223T, 2020MNRAS.491.2318T, 2020MNRAS.tmp.1883K}.

The major systematic uncertainty in tSZ measurements has been the contamination of the cosmic infrared background (CIB) sourced by thermal dust emission of unresolved galaxies \citep{2005ARA&A..43..727L}. Since the amplitude and observer-frame spectrum of the CIB are strongly redshift dependent, so is its impact on the tSZ measurement \citep{2016A&A...594A..23P}. This limits the extent to which one can clean the CIB from any map-level reconstruction of the tSZ field as used in the previous work. 

To best handle the redshift-dependent CIB impact and explore tSZ constraints up to the highest possible redshifts, in this paper we present a full multichannel redshift tomography for the far-infrared to microwave extragalactic background light (EBL)\footnote{In this work we refer the term EBL to the generic body of the extragalactic radiation field that traces the underlying matter density field. It includes the CIB dust emission and the tSZ effect but not the CMB.}. We start by performing clustering-based redshift measurements separately for each of eight intensity maps of the {\it Planck} and the Infrared Astronomical Satellite ({\it IRAS}) missions. This allows us to obtain, at each redshift bin, a deprojected snapshot of the cosmic spectral energy distribution (SED). Using its distinct spectral feature, the tSZ signal can then be separated from the evolving CIB. Our tSZ measurements directly constrain the halo bias-weighted mean electron pressure $\langle bP_{\rm e} \rangle$ over cosmic time \citep{2017MNRAS.467.2315V}. By combining the $\langle bP_{\rm e} \rangle$ measurements and a halo model-based bias correction, we probe the density-weighted mean temperature of electrons $\overbar{T}_{\rm{e}}$. We further constrain, for the first time, the comoving mean thermal energy density $\rho_{\rm th}$ and the corresponding density parameter $\Omega_{\rm th}$, which represents a fundamental summary statistic of the cosmic thermal history. 

This paper is organized as follows. In Section~\ref{sec:analysis} we describe the data products and introduce the clustering redshift formalism.
In Section~\ref{sec:halo-model} we describe a halo model for interpreting our results. In Section~\ref{sec:tSZ_background} we present the tomographically measured tSZ amplitudes, the derived tSZ mass bias parameter, and the mean pressure constraints. In Section~\ref{sec:discussion_Omega_th} we interpret our tSZ measurements in terms of the cosmic thermal history before concluding the work in 
Section~\ref{sec:conclusion}. Throughout the paper, we assume a $\rm \Lambda CDM$ cosmology with the Planck-2018 ``$\rm TT,\ TE,\ EE+lowE+lensing$'' parameters in Table~1 of \cite{2020A&A...641A...6P}: ($h$, $\Omega_{\rm c}h^2$, $\Omega_{\rm b}h^2$, $A_{\rm s}$, $n_{\rm s}$) = (0.6737, 0.1198, 0.02233, $2.097\times10^{-9}$, 0.9652) with the minimal sum of the neutrino masses of 0.06 eV. The present-day matter density parameter is $\Omega_{\rm m}=\Omega_{\rm c}+\Omega_{\rm b}+\Omega_{\nu}=0.3146$, which sets the cosmological constant parameter $\Omega_\Lambda = 1 - \Omega_{\rm m}$.

\section{Thermal SZ Tomography}
\label{sec:y}
The amplitude of the tSZ effect is quantified by the Compton $y$ parameter \citep{SZ_1972}:
\begin{equation}
y(\hat{\phi}) = \frac{\sigma_\textrm{T}}{m_\textrm{e}\, c^2} \int \frac{\textrm{d}\chi}{1+z}\, P_\textrm{e}(\chi\,\hat{\phi}), 
\label{eq:y}
\end{equation}
where $\hat{\phi}$ is the sky direction vector, $\sigma_\textrm{T}$ the Thomson scattering cross section, $m_\textrm{e}$ the electron mass, $c$ the speed of light, $P_\textrm{e}=n_\textrm{e}\, k_\textrm{B}\, T_\textrm{e}$ the electron pressure with $T_\textrm{e}$ and $n_\textrm{e}$ being the electron temperature and (proper) density, respectively, $k_\textrm{B}$ the Boltzmann constant, and $\chi = \chi(z)$ the comoving radial distance to redshift $z$. The integral goes from zero to that at the surface of the last scattering. The total thermal gas pressure is given by $P_{\rm th} = [(8-5Y)/(4-2Y)]\,P_\textrm{e}$, assuming that the gas is fully ionized and $Y$ is the primordial helium mass fraction. For $Y=0.24$, $P_{\rm th} = 1.932\,P_\textrm{e}$.

The tSZ effect gives rise to a unique spectral distortion of the CMB. We relate the observed distortion in specific intensity as a function of frequency to the Compton $y$ parameter as \citep{2002ARA&A..40..643C}:
\begin{equation}
\Delta I_{\nu}(\hat{\phi},\, \nu) = g(x)\, I_{\nu,\,0}\, y(\hat{\phi}),
\label{eq:delta_I_nu_given_y}
\end{equation}
where $I_{\nu,\,0} = 2\,(k_\textrm{B}\,\textrm{T}_\textrm{CMB})^3/(hc)^2$ with $\textrm{T}_\textrm{CMB} = 2.725$\,K. The frequency dependence is given by
\begin{equation}
g(x) = \frac{x^4\, e^x}{(e^x-1)^2}\, \Bigg( x\, \frac{e^x+1}{e^x-1}-4 \Bigg),
\label{eq:g_x}
\end{equation}
where $x \equiv h\,\nu/(k_\textrm{B}\,\textrm{T}_\textrm{CMB})$. Here we neglect the relativistic correction as it is too small to impact our measurements. Given a physical $y>0$, the CMB spectrum receives an decrement (increment) below (above) $218$~GHz, with no effect at $218$~GHz. 

In this paper, we focus on the quantity $\textrm{d}\langle y\rangle/\textrm{d}z$, where $\langle ... \rangle$ denotes an ensemble average. Hereafter we simply write $\textrm{d}y/\textrm{d}z$. This tomographic quantity for the cosmic tSZ history is of utmost importance in our study, as it undoes the line-of-sight integral in Equation~\ref{eq:y} and scales directly with the (deprojected) cosmic mean electron pressure $\langle P_{\rm e}\rangle$. As we show in Section~\ref{sec:clustering-redshift}, the clustering redshift technique provides us with an only slightly modulated observable, $\textrm{d}y/\textrm{d}z \times b_y$, where $b_y$ is the $y$-weighted, large-scale halo clustering bias $b$. We can thus directly constrain the cosmic mean bias-weighted electron pressure \citep{2017MNRAS.467.2315V}:

\begin{equation}
\langle b P_{\rm e}\rangle = b_y \langle P_{\rm e}\rangle = \frac{m_\textrm{e}\, c^2\, (1+z)}{\sigma_\textrm{T}}\, \frac{\textrm{d}z}{\textrm{d}\chi}\, \frac{\textrm{d}y}{\textrm{d}z}\,b_y\,,
\label{eq:bPe}
\end{equation}
from our clustering-based tSZ tomography measurements.
Given the basis of $\rm \Lambda CDM$ structure formation, $b_y$ can be modeled robustly in a halo model (Section~\ref{sec:halo-model}). The key quantity $\textrm{d}y/\textrm{d}z$ can thus be measured in a nearly model-independent and empirical manner.

As the main motivation of this work is to probe the thermal history of the universe using tSZ tomography, we define the mean comoving thermal energy density,
\begin{equation}
\rho_\textrm{th} \equiv \frac{\langle P_\textrm{th} \rangle}{(1+z)^3} = \frac{m_\textrm{e}\, c^2\, (8-5Y)}{\sigma_\textrm{T}\, (4-2Y)\, (1+z)^2}\, \frac{\textrm{d}z}{\textrm{d}\chi}\, \frac{\textrm{d}y}{\textrm{d}z}\,,
\label{eq:rho_th}
\end{equation}
where ${\rm d}z/{\rm d}\chi=H(z)/c$, as well as the corresponding density parameter,
\begin{equation}
\Omega_\textrm{th}(z) = \frac{\rho_\textrm{th}(z)}{\rho_\textrm{crit}}\,,
\label{eq:omega_th}
\end{equation}
where $\rho_\textrm{crit} = 1.054\times 10^4~h^2~{\rm eV~cm^{-3}}$ is the critical density of the universe at $z=0$ expressed in energy units. 

We can also express $\textrm{d} y/\textrm{d}z$ in terms of the density-weighted mean temperature of electrons in the universe \citep{1999ApJ...514....1C,2000PhRvD..61l3001R}, defined by
\begin{equation}
\overbar{T}_e \equiv \frac{\langle n_e\,T_e\rangle}{\langle n_e \rangle} = \frac{2\, m_\textrm{H}\, m_\textrm{e}\, c^2}{\rho_\textrm{crit}\, \Omega_{\rm b}\, \sigma_\textrm{T}\, k_{\rm B}\, (2-Y)\, (1+z)^2}\, \frac{\textrm{d}z}{\textrm{d}\chi}\, \frac{\textrm{d}y}{\textrm{d}z}\,,
\label{eq:T_bar}
\end{equation}
where $m_\textrm{H}$ is the mass of a hydrogen atom. Here we have assumed that the gas retains the primordial chemical abundance and is fully ionized.

The evolutions of these cosmic mean thermal properties, $\langle{P_{\rm{e}}}\rangle$, $\Omega_\textrm{th}$, and $\overbar{T}_{\rm{e}}$, are all connected to the tSZ amplitude $\textrm{d}y/\textrm{d}z$, and are driven mainly by the build-up of cosmic structures. Tomographic observations of the tSZ background thus provide a powerful way to probe the cosmic structure formation.

\section{Analysis}
\label{sec:analysis}
In this paper we measure $\textrm{d}y/\textrm{d}z$ as a function of redshift in two ways:
\begin{itemize}
    \item Clustering-based redshift estimations using the Compton $y$ maps generated by \cite{planck2016_sz_map}.
    However, we find clear evidence for the CIB contamination in these maps; thus, we consider this as a supplementary analysis that will only be used to check consistency.
    \item Redshift-dependent component separation. In this approach, we start by deprojecting each of eight intensity maps $I_{\nu_i}$ using the clustering redshift technique and obtain the EBL SED containing both the tSZ and CIB redshift by redshift. To extract $\textrm{d}y/\textrm{d}z$, we simultaneously fit SEDs of the tSZ (Equation~\ref{eq:delta_I_nu_given_y}) and redshift-evolving CIB. 
\end{itemize}
In this section we describe the data products used in this paper, our data processing, and the clustering redshift estimation technique.

\subsection{Data}
\label{sec:data}
\subsubsection{Planck and IRAS Intensity Maps}
\label{sec:data-Planck-intensity}
To cover the spectral windows of the tSZ effect and the CIB, we use eight full-sky intensify maps. These include six channels at 100, 143, 217, 353, 545, and 857\,GHz from the {\it Planck} satellite High Frequency Instrument \citep[HFI;][]{Planck_2016_HFI} and the 100 and 60\,$\mu m$ channels (3000 and 5000\,GHz, respectively) form the Infrared Astronomical Satellite ({\it IRAS}) reprocessed by \cite{2005ApJS..157..302M}. The maps span a moderate range of spatial resolution, with the beam FWHMs of $10'$, $7.1'$. $5.5'$, $5'$, $5'$, $5'$, $4.3'$, and $4'$ from low to high frequencies. We apply a common spatial sampling scheme using HEALPix \citep{Gorski_2015} with an $N_{\rm side}$ of 2048. We convert the map units from thermodynamic temperature to specific intensity in MJy\ sr$^{-1}$ using the {\it IRAS} convention assuming $\nu\,I_\nu = \rm constant$ within each band. A ``color correction'' \citep{2014A&A...571A...9P} is used to account for the likely departure of the true spectrum from this assumption when we perform the multichannel SED fitting in Section~\ref{sec:results-SED}.

The sky intensities in these channel maps are dominated by four astrophysical components: 
(1) Galactic foreground, (2) the primary CMB, (3) the tSZ effect, and (4) the CIB. In our clustering-based redshift tomography, only the spatial fluctuations of the latter two components contribute to the signals. Galactic foreground gradients on large scales dominate the systematic errors, and both the Galactic and the primary CMB anisotropies on small scales dominate the statistical errors. To obtain robust results under the presence of the these fore- and backgrounds, we clean the maps with the following procedure.

We apply a common mask to all channels, which is the union of a large-scale Galactic mask and a small-scale bright-point-source mask\footnote{The masks are available at https://zenodo.org/record/4012781.}. At large scales, we mask $60\%$ of the sky with the highest intensities in the {\it Planck} 143 GHz channel (smoothed over 5$^{\circ}$), which lie mainly at low Galactic latitudes. The choice of 143 GHz as the Galactic mask reference is to optimize the signal-to-noise ratio (S/N) of the final tSZ extraction: 143 GHz is near the peak frequency of the tSZ decrement and also traces the broad thermal dust continuum emission from Galactic dust. The choice of the $60\%$ masked fraction is based on maximizing the S/N of the tSZ measurements while avoiding significant bias by the Galactic foreground. At small scales, in all our $Planck$ and {\it IRAS} maps we mask a common set of bright point sources that are individually detected in at least one {\it Planck} HFI channel. We ensure that the per-source masking radius is at least $10'$ to match the beam of our lowest-resolution map.

We use a template-based cleaning method to reduce the small-scale fluctuations of the primary CMB and the diffuse Galactic dust emission while keeping the signals from extragalactic structures largely unchanged. For each {\it Planck} HFI intensity map from 100 to 353 GHz, we remove the beam-matched primary CMB contribution using the CMB map estimated in \cite{Bobin_2016} based on {\it Planck} and Wilkinson Microwave Anisotropy Probe data \citep{Bennett_2013, Planck_2016_LFI, Planck_2016_HFI}\footnote{The CMB-subtracted channel maps are available at https://zenodo.org/record/4012781.}. A $\rm T_{CMB}=2.725\,K$ blackbody spectrum is assumed to convert the CMB map in $\rm \Delta K_{CMB}$ units into intensity units at each frequency channel. The \cite{Bobin_2016} map is preferred over other CMB maps because it is shown to contain no noticeable tSZ residual, which minimizes the risk of altering the signals that we wish to extract. 

We also apply a template-based cleaning method to reduce the impact of Galactic thermal dust emission in all of the {\it Planck} and {\it IRAS} channels we use. Similar to the CMB cleaning, the Galactic dust cleaning should not alter the extragalactic signals, while most broadband-based dust maps are found to contain detectable levels of CIB \citep{Chiang_Menard_2019}. We thus use HI as a proxy for dust for which spectroscopic constraints are available through the 21~cm emission. We take the map of HI column density (within $90\ \rm km\ s^{-1}$ from the local standard of rest) from \cite{Lenz_2017}, constructed using the data from \cite{HI4PI_2016}. Different from the direct CMB subtraction above, the HI template is only an indirect dust tracer. Dust and gas are tightly correlated only up to about $15^{\circ}$ scale \citep{1998ApJ...500..525S} and the HI-to-dust-emission conversion factor could depend on frequency. We take an empirical approach in obtaining these conversion factors on the scales appropriate to our clustering redshift measurements. We first beam-match the CMB-removed channel maps to that of the HI map ($\rm FWHM=16.1'$) and perform a $15^{\circ}$ high-pass filtering for all maps. We then perform a linear regression between the HI column density and each channel intensity using the residual fluctuations in the unmasked area of the high-pass filtered maps. The HI-predicted fluctuations in each frequency are then subtracted from the unsmoothed, CMB-removed channel maps.

The end product is a set of eight masked channel intensity maps with, on average, zero CMB contribution over all scales and reduced Galactic foregrounds on scales ranging from $16'$ to $15^{\circ}$.

\subsubsection{Planck y Maps for Comparison}
\label{sec:data-Planck-y}
For comparison, we also use two full-sky maps of the Compton $y$ parameter from \cite{planck2016_sz_map} and extract the cosmic tSZ history via clustering-based redshift estimations. These maps are constructed using two implementations of the internal linear combination (ILC) method \citep{bennett/etal:2003,tegmark/etal:2003,eriksen/etal:2004,hinshaw/etal:2007}: the Needlet ILC \citep[NILC;][]{delabrouille/etal:2009,Remazeilles_2011} and the Modified ILC Algorithm \citep[MILCA;][]{Hurier_2013}. These $y$ maps are based on the same {\it Planck} channel intensity maps that we use (while 30, 44, and 70~GHz are included additionally in building the NILC $y$ map).  The $y$ maps have a beam of $10'$ and are pixelized using the HEALPix scheme with an $N_{\rm side}$ of 2048. We apply the same Galactic and point-source masks as those used for the channel intensity maps. However, we do not apply the template-based CMB and Galactic foreground cleaning here as the attempt at component separation has been made via the NILC and MILCA algorithms.

\subsubsection{SDSS Large-scale Structure Reference}
We compile a sample of two million spectroscopic sources as the reference sample for our clustering-based redshift analyses. This compilation consists of galaxies and quasars with secured spectroscopic redshifts up to $z\sim3$ from seven catalogs released as part of the Sloan Digital Sky Survey (SDSS), Baryon Oscillation Spectroscopic Survey (BOSS), and Extended Baryon Oscillation Spectroscopic Survey (eBOSS) up to the public Data Release (DR) 14. At $z\lesssim 0.15$, we rely on the large-scale structure catalog in \cite{2005AJ....129.2562B} based on the flux-limited SDSS MAIN galaxy selection \citep{2002AJ....124.1810S}. At $0.15\lesssim z \lesssim 0.8$, we use three large-scale structure catalogs of luminous red galaxies \citep[LRGs; ][]{2001AJ....122.2267E}, which include the LOWZ and CMASS samples \citep{2016MNRAS.455.1553R} from BOSS and the DR14 LRG sample from eBOSS \citep{2018ApJ...863..110B}. We also use three quasar (QSO) catalogs: SDSS QSO \citep{2010AJ....139.2360S}, BOSS QSO \citep{2017A&A...597A..79P}, and the eBOSS QSO sample \citep[DR14;][]{2018MNRAS.473.4773A}, which allows us to extend the redshift coverage up to $z\sim3$. The sky footprints vary moderately between the SDSS and BOSS catalogs and are significantly smaller for the eBOSS catalogs. As a result, the effective sky area used is redshift dependent as different catalogs cover different redshift ranges. We compile a set of small-scale rejection masks (bright stars, bad imaging tracks, and small regions without spectroscopic coverage) provided by these surveys and generate a joint rejection mask by taking the union\footnote{The masks are available at https://zenodo.org/record/4012781.}.

Table~\ref{tab:spec-z-ref} summarizes the basic parameters of the spectroscopic reference sources that we use after applying the joint SDSS rejection mask, the 60\% Galactic mask, and the infrared point-source mask described in \S~\ref{sec:data-Planck-intensity}.

\renewcommand{\arraystretch}{1.2}

\begin{table}[t]
\centering
\caption{\label{table:reference-sample}}
Spectroscopic Redshift Reference\\
\vspace{0.2cm}
\begin{tabular}{@{}cccccccc@{}}
\hline
\hline
survey & sample    & A$^{a}$ & $\#^b$ & $f_{\rm N}$$^{c}$ & redshift$^{d}$         & ref.     \\ \hline
SDSS   & MAIN          &  7342  & 440220                     & 94\%                  & $0.09^{+0.06}_{-0.04}$ & $^{e}$,$^{f}$ \\
BOSS   & LOWZ      &  7554    & 330644                    & 82\%                  & $0.31^{+0.09}_{-0.15}$ & $^{g}$         \\
BOSS   & CMASS     &  8307  & 616571                    & 85\%                  & $0.54^{+0.09}_{-0.07}$ & $^{g}$         \\
eBOSS  & LRG           &  2210  & 107365                    & 66\%                  & $0.68^{+0.13}_{-0.07}$ & $^{h}$         \\
SDSS   & QSO           &  5806  & 48602                     & 99\%                  & $1.48^{+0.87}_{-0.80}$ & $^{i}$         \\
BOSS   & QSO           &  8146  & 90080                     & 87\%                  & $2.43^{+0.51}_{-0.23}$ & $^{j}$         \\
eBOSS  & QSO           &  2258    & 14,323                    & 69\%                  & $1.60^{+0.62}_{-0.57}$ & $^{k}$        \vspace{0.075cm} \\  
\hline
\multicolumn{2}{c}{combined sample}    &  8641  & 1782805                    & 85\%                  & $0.48^{+0.31}_{-0.37}$ &          \vspace{0.075cm} \\
\hline
\end{tabular}
\begin{flushleft}
$^{a}$ Effective unmasked sky area in deg$^2$;\\
$^{b}$ Number of sources in the effective area;\\
$^{c}$ Fraction of the sources in the Northern Galactic Hemisphere;\\
$^{d}$ Median and 68\% CL range of the redshift distribution;\\
$^{e}$ \cite{2002AJ....124.1810S}; $^{f}$ \cite{2005AJ....129.2562B}; $^{g}$ \cite{2016MNRAS.455.1553R}; \\ $^{h}$ \cite{2018ApJ...863..110B}; $^{i}$ \cite{2010AJ....139.2360S}; \\ $^{j}$ \cite{2017A&A...597A..79P}; $^{k}$ \cite{2018MNRAS.473.4773A}.
\end{flushleft}
\label{tab:spec-z-ref}
\end{table}

\subsection{Clustering Redshifts}
\label{sec:clustering-redshift}

We apply the clustering-based or cross-correlation-based redshift estimation following \cite{Menard_2013} for both the channel intensity and {\it Planck}'s Compton $y$ maps. The technique is based on the simple ansatz that all matter tracers appear spatially clustered on the sky if their redshift distributions overlap. One can thus take a ``reference'' sample of galaxies or quasars with known redshifts and propagate the redshift information into a ``test'' dataset with an unknown redshift distribution. This is achieved by cross-correlating the reference and test samples as a function of redshift of the former. 

The first applications to large datasets are done in the regime of discrete objects as the test sample to estimate source redshift distributions in photometric surveys \citep[e.g.,][]{2015MNRAS.447.3500R, 2016MNRAS.457.3912R, 2016MNRAS.460..163R, 2016MNRAS.462.1683S, 2017MNRAS.465.1454H, 2018MNRAS.477.2196D, 2018MNRAS.477.1664G, 2020MNRAS.496.2262K, 2020JCAP...05..047K}. It is straightforward to generalize the method and redshift inference to diffuse fields simply by replacing the notion of objects with pixels on a set of predetermined grids \citep{Schmidt_2015, Chiang_Menard_2019, Chiang_2019}. Our implementation in this work is largely based on that in \cite{Chiang_2019}; here we briefly describe the key steps and minor modifications made for the tSZ tomography.

We denote the overdensity field of the 2D test intensity map as $\rm T(\hat{\phi})$ (where $\textrm{T} = I_{\nu_i}$ in the multichannel analysis and $\textrm{T} = y$ for the direct $y$ map tomography) and that for the 3D reference sources as $\rm R(\hat{\phi},\, z)$. As both are, in general, biased tracers of the underlying matter density field with the angular correlation function $\overline{w}_{\rm DM}$ at some effective scale, we can write down a linear expression for their cross-correlation amplitudes $\overline{w}_{\rm TR}$ evaluated in bins of the reference redshift $z_i$ as
\begin{eqnarray}
\overline{w}_{\rm TR}(z_i) = 
\frac{\rm{dT}}{\textrm{d}z}(z_i)\, b_{\rm T}(z_i) \, b_{\rm R}(z_i)\,\overline{w}_{\rm DM}(z_i)\;,
\label{eq:wbar_to_dJdz}
\end{eqnarray}
where $b_{\rm T}$ and $b_{\rm R}$ are the effective linear-clustering bias for the test and reference data at the scales considered. The target quantity $\rm{dT}/\textrm{d}z$, that is, the redshift derivative of the intensity of the test field, appears as a normalization factor. In this expression, the left-hand side is our primary observable. On the right-hand side, the dark matter clustering $\overline{w}_{\rm DM}$ can be calculated once a cosmological model is assumed, and the bias of the reference sample $b_{\rm R}$ can be measured empirically. The clustering redshift estimation thus empirically constrains the product $\textrm{dT}/\textrm{d}z(z_i)\, b_{\rm T}(z_i)$. 

Equation~\ref{eq:wbar_to_dJdz} is exact on large scales, where clustering amplitudes can be described by the ``two-halo'' term contribution in the standard halo model formalism \citep[see][for a review]{2002PhR...372....1C}. We calculate $\overline{w}_{\rm DM}$, the effective dark matter clustering amplitude at each redshift bin based on the dark matter angular correlation function $w_{\rm DM}(\theta)$. We calculate $w_{\rm DM}(\theta)$ using Equation~10 in \cite{Chiang_2019} based on the nonlinear matter power spectra generated by the {\sf CLASS} code \citep{blas/lesgourgues/tram:2011, 2011JCAP...09..032L}, which uses the ``Halofit'' formalism \citep{smith/etal:2003} with parameters from \cite{takahashi/etal:2012}. To obtain $b_{\rm R}$ in Equation~\ref{eq:wbar_to_dJdz}, we measure the auto correlations of the reference sample and solve $\overline{w}_{\rm RR} = b_{\rm R}^2\,\overline{w}_{\rm DM}$ for $b_{\rm R}$ at each redshift bin. The measured $b_{\rm R}$ is similar to that shown in Figure~12 in \cite{Chiang_Menard_2019} and is measured with a percent-level precision at $z<0.8$. For Compton $y$ as the test sample, the bias $b_{\rm T} = b_y$ is the effective clustering bias for the $y$ field with respect to the matter density field. It is generally not directly measured but can be robustly modeled using a halo-model-based approach, as will be shown in Section~\ref{sec:halo-model}. 

Having laid out the basic principles, we now describe a few technical details. Equation~\ref{eq:wbar_to_dJdz} is, in general, valid on all linear-clustering scales, with the target $\rm{dT}/\textrm{d}z$ being, by definition, scale-independent. We therefore need to specify the scheme at which signals from a range of scales are combined. This defines the overhead bar notation in $\overline{w}_{\rm TR}$ and $\overline{w}_{\rm DM}$. Given a two-point correlation function $w(\theta)$, we perform an angular integral following \cite{Menard_2013}:
\begin{eqnarray}
\overline{w} = \int_{\theta_{\rm min}}^{\theta_{\rm max}} W(\theta)\,w(\theta)\, \textrm{d}\theta\;,
\label{eq:wbar}
\end{eqnarray}
with $W(\theta)$ being an arbitrary weight function, and $\theta_{\rm min}$ and $\theta_{\rm max}$ being the minimum and maximum scales considered for the measurements. For an optimal estimator, we take the expected signal as the weight, $W(\theta) \propto w_{\rm DM}(\theta,z)$, which is the dark matter two-point autocorrelation function. 

We use redshift-dependent ($\theta_{\rm min}$, $\theta_{\rm max}$) corresponding to fixed physical separations ($r_{\rm p,\, min}$, $r_{\rm p,\, max}$). We set $r_{\rm p,\, max} = 8$~Mpc for all our analyses; at $z>0.6$, this corresponds to an angular size of about 16', where the systematics of large-scale zero-point fluctuations are negligible. The $\theta_{\rm min}$, on the other hand, needs to be chosen to meet more physical and experimental constraints. As the small-scale clustering is stronger, it is beneficial to integrate the signal down to a smaller $r_{\rm p,\, min}$ as long as the linear Equation~\ref{eq:wbar_to_dJdz} is still valid. \cite{2017MNRAS.467.2315V} shows that the tSZ--galaxy group cross-correlation function is dominated by the two-halo term at $\gtrsim 2$~Mpc, which sets a minimum linear scale for our analysis. Some of the {\it Planck} maps we use, however, have a beam large enough to affect the clustering measurements at this scale at high redshifts. For these considerations, we set $r_{\rm p,\, min} = 3$~Mpc for 100 and 143~GHz and {\it Planck}'s Compton $y$ maps with relatively large beams, and $r_{\rm p,\, min} = 2$~Mpc for the rest, the slightly higher resolution maps at higher frequencies. As our multichannel tSZ tomography relies mainly on the signals of the tSZ decrements at 100 and 143 GHz, the effective scale for our tSZ extraction is 3--8 Mpc (physical). In Appendix~\ref{appendix:1_halo_impact} we investigate the potential bias of clustering redshift estimation from the one-halo term on small scales, and we show that the impact is negligible. 

As mentioned in Section~\ref{sec:data-Planck-intensity}, we perform template-based cleaning to reduce the Galactic foreground emission and the primary CMB on the test intensity maps. When carrying out the clustering redshift measurements, we additionally perform a high-pass filtering with a Gaussian kernel to remove any remaining large-scale fluctuations that are likely dominated by foregrounds. To avoid altering signals that we wish to extract at the maximum scales used ($\theta_{\rm max}$, which is larger at lower redshifts), we use a more relaxed Gaussian filtering scale of $\rm FWHM=4$~deg at $z<0.25$ and a more aggressive $\rm FWHM=2$~deg at $z>0.25$.

When modeling the cross-correlation amplitudes, we take into account the effects of the map resolution and data processing. Both the instrument beams and the high-pass filtering that we apply reduce the clustering amplitudes $\overline{w}_{\rm TR}$ in a map-, scale-, and redshift-dependent way. We include these two effectively small- and large-scale  window functions on the right-hand side of Equation~\ref{eq:wbar_to_dJdz} when modeling our estimator. Specifically, we multiply the window functions and the theoretical matter power spectrum in Fourier space to generate modified $w_{\rm DM}(\theta)$ and $\overline{w}_{\rm DM}$ taking into account the loss of clustering power. We have chosen ($r_{\rm p,\, min}$, $r_{\rm p,\, max}$) and the filtering scales described above to balance the foreground mitigation, instrument beams, and the ability to model linear clustering, while minimizing window function corrections. 

In our $\overline{w}_{\rm TR}$ measurements, we estimate the error bars empirically via a resampling technique. A jackknife approach is not preferred because in our case the sky area depends on redshift (Table~\ref{tab:spec-z-ref}) and the foreground varies strongly with frequency; the jackknife errors would thus depend strongly on the scheme at which the jackknife regions are divided. We thus use the bootstrapping approach with a modification to the standard implementation. The basic assumption in bootstrapping is that the resampled data units can be treated as independent. The errors would thus be underestimated if the resampling is done for reference sources that are clustered. Similarly, one needs to be cautious if the resampling is done on the test maps for which more complex spatial correlations are present. In this work we spatially group the reference sources using a redshift-dependent grouping length such that the intergroup clustering amplitude is much smaller than unity. The bootstrap resampling is then done on groups of reference objects. We fix the exact set of resampled reference groups when estimating the errors of $\overline{w}_{\rm TR}$ with different channel intensity maps as the test set. This allows us to empirically estimate the frequency covariance matrix at each redshift bin.

We expect negligible covariance for measurements at different redshift bins as our reference sample is spectroscopic, and the redshift bin width is much larger than that corresponds to the typical correlation length in the cosmic web ($\lesssim 10$~Mpc). This is in contrast to the previous tSZ measurements using photometric galaxy catalogs \citep{2019PhRvD.100f3519P,2020MNRAS.491.5464K}.

The clustering redshift measurements presented in this paper can be reproduced using the {\sf Tomographer} (Chiang et al., in prep), a web-based platform at \url{http://tomographer.org}.

\section{Halo Model}
\label{sec:halo-model}
To interpret our tomographic measurements of the bias-weighted tSZ amplitudes $\rm{dy}/\textrm{d}z\times b_{y}$, we use a halo model as presented in \citet{2018MNRAS.480.3928M,2020PASJ...72...26M}\footnote{Codes are available in \url{https://github.com/ryumakiya/pysz}.}, which is based on \cite{1999ApJ...526L...1K}, \cite{2002MNRAS.336.1256K}, and \cite{2018MNRAS.477.4957B}. In this formalism, the mean tSZ signal is originated from the hot gas in a population of dark matter halos, especially massive clusters with high virial temperatures. In the model we neglect the contribution from the diffuse intergalactic medium, which is two orders of magnitudes below the halo contribution \citep{2015PhRvL.115z1301H}. 

The redshift derivative of the mean Compton $y$, ${\rm d}y/{\rm d}z$ (corresponds to ${\rm dT}/{\rm d}z$ in Equation~\ref{eq:wbar_to_dJdz}), is given by
\begin{equation}
\frac{{\rm d}y}{{\rm d}z} = \frac{{\rm d}V}{{\rm d}z{\rm d}\Omega} \int_{M_{\rm min}}^{M_{\rm max}} {\rm d}M \frac{{\rm d}n}{{\rm d}M} \tilde{y}_{0}(M,z),
\label{eq:dYdz}
\end{equation}
where $\tilde{y}_0$ is the total Compton $y$ contributed from a halo of mass $M$ at redshift $z$. For the dark matter halo mass function, $\textrm{d}n/\textrm{d}M$, we use that given in \cite{tinker/etal:2008}. 
We set a mass integration range of $10^{11} \ h^{-1}\ \textrm{M}_{\odot} < M_{\rm 500} < 5\times 10^{15}\ h^{-1}\ \textrm{M}_{\odot}$ where $M_{\rm 500}$ is the total mass enclosed within $r_{\rm 500}$, the radius within which the mean matter overdensity is 500 times of the critical density of the universe. This allows the integral to converge to better than (0.1\%, 1\%, 3\%, 10\%) at $z=($0, 1, 2, 3); this is sufficient because the uncertainty in the halo mass function at each corresponding redshift is considerably larger. The mass functions of \cite{tinker/etal:2008} are given for halo masses defined at given overdensities with respect to the mean mass density rather than the critical density. We thus interpolate the parameters at various mean mass overdensities to obtain the mass functions for $M_{500}=M_{\rm 500c}$.

In the calculation of $dn/dM$ we include the effect of massive neutrinos by following the so-called ``CDM prescription'' (\citealt{ichiki/takada:2012,costanzi/etal:2013,castorina/etal:2014,villaescusa-Navarro/etal:2014}). The basic idea is to remove the contribution of neutrinos from the mass of collapsed halos when computing statistics of halos, as neutrinos stream out of them. For a fixed sum of the neutrino mass of $\rm 0.06\ eV$, the effect on tSZ amplitudes is about $2\%$. We refer the readers to \cite{2020MNRAS.497.1332B} for more details.

Given the definition of Compton $y$ in Equation~\ref{eq:y}, $\tilde{y}_{0}$ can be written as
\begin{equation}
\tilde{y}_{0}(M,z) = 
\frac{r_{\rm 500}^3}{D_{\rm A}^2}
\frac{\sigma_{\rm T}}{m_{\rm e} c^2}
\int^{r_{\rm max}}_{0}{\rm d}r\;4\pi r^2 P_{\rm e}(r| M, z),
\label{eq:y_0}
\end{equation}
where $P_{\rm e}$ is the radial distribution of the electron pressure, and $D_{\rm A}$ is the proper angular diameter distance.
The integral is performed out to $r_{\rm max}/r_{\rm 500} = 6$, beyond which the pressure falls rapidly \citep{bryan/norman:1998,shi:2016}.

For the electron pressure profile, we use that given by \cite{arnaud/etal:2010}:
\begin{eqnarray}
\label{eq:pe}
P_{\rm e}(x) &=& 1.65\,(h/0.7)^2 {\rm \;eV\;cm^{-3}} \nonumber \\ 
 &\times& E^{8/3}(z) \left[ \frac{M_{\rm 500}}{3 \times 10^{14}\,(0.7/h)\,{\rm M}_{\odot}}\right]^{2/3+\alpha_p}p(x),
\end{eqnarray}
where $x \equiv r/r_{\rm 500}$ and $E(z) \equiv H(z)/H_0$.
The parameter $\alpha_p$ quantifies the deviation from the self-similar behavior, with $\alpha_p = 0$ being exactly self-similar. We use a fixed value of $\alpha_p = 0.12$ based on the best fit in \cite{arnaud/etal:2010} using a sample of X-ray clusters. For the self-similar part of the pressure profile, $p(x)$, we use the generalized \citet*{1997ApJ...490..493N} profile defined by \cite{nagai/etal:2007}:
\begin{equation}
p(x) \equiv \frac{P_{0}\,(0.7/h)^{3/2}}{(c_{\rm 500}\,x)^{\gamma}\,
[1+(c_{\rm 500}\,x)^{\alpha}]^{(\beta-\gamma)/\alpha}}\,.
\label{eq:px}
\end{equation}
For the parameters in $p$ we use those in \cite{planck_inter_v:2013}: $(P_{0},\,c_{\rm 500},\,\alpha,\,\beta,\,\gamma) = (6.41,$ 1.81, 1.33, 4.13, 0.31), which are obtained via jointly fitting the stacked tSZ and X-ray profiles for a sample of nearby massive clusters.

The mass--pressure relation in Equation~\ref{eq:pe} is calibrated empirically by combining X-ray and $\it Planck$ tSZ observations assuming hydrostatic equilibrium between gravity and the thermal pressure gradient. The relation could thus be biased by the presence of non-thermal pressure support or unaccounted observational or calibration bias. To account for the unknown absolute mass calibration, we introduce a ``mass bias'' parameter 
\begin{equation}
B = M_{\rm 500,\ true} / M_{\rm 500,\ empirical}, 
\label{eq:mass_bias_B}
\end{equation}
where $M_{\rm 500,\ true}$ and $M_{\rm 500,\ empirical}$ are the true mass in the halo mass function and the empirically calibrated mass used in Equation~\ref{eq:pe}, respectively. We do not call it a ``hydrostatic bias'' as effects beyond the assumption of hydrostatic equilibrium could enter.

To include the mass bias in the model, we rescale the $M_{\rm 500}$ and $r_{\rm 500}$ in Equation~\ref{eq:pe} to $M_{\rm 500}/B$ and $r_{\rm 500}/B^{1/3}$, respectively. The mass bias is the only free parameter in the halo model in determining the tSZ amplitudes with a simple scaling: $\textrm{d}y/\textrm{d}z\propto B^{-5/3-\alpha_p}$. In this work, we allow for redshift dependence of $B$, while we assume it to be mass independent, as our measurements of the mean tSZ background would not be sensitive to the mass dependence.

\begin{figure}[t!]
    \begin{center}
         \includegraphics[width=0.46\textwidth]{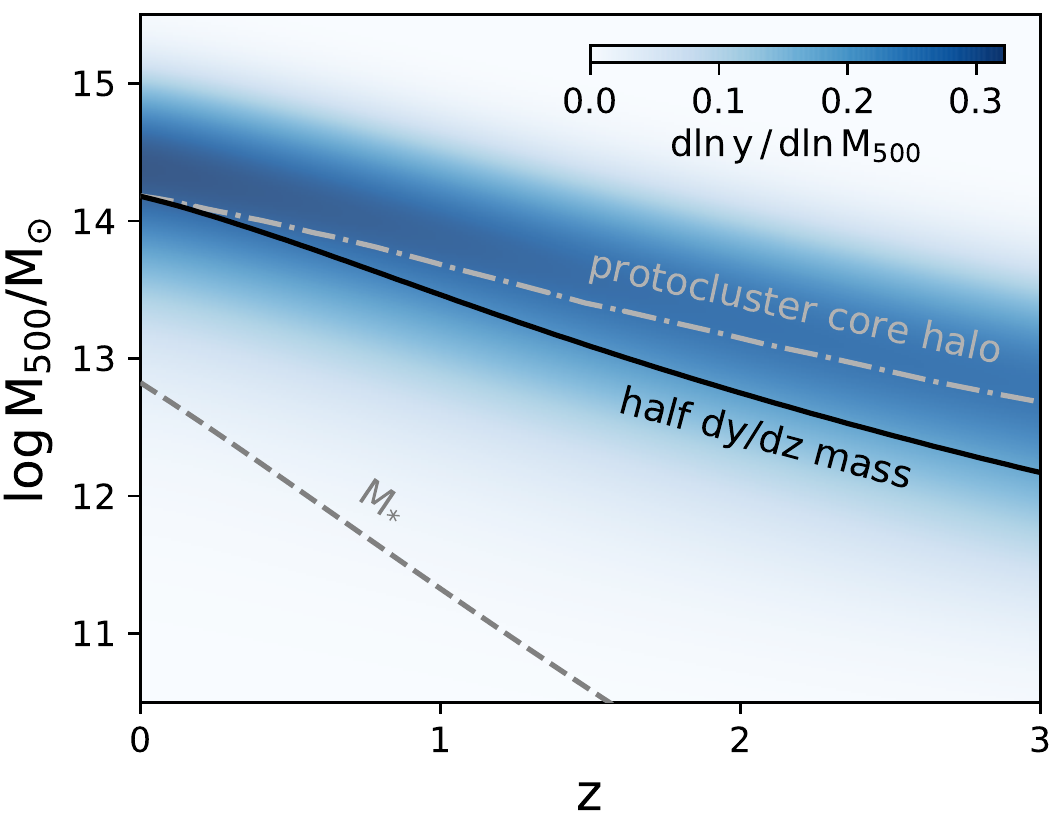}
    \end{center}
    \caption{Differential contribution to the deprojected tSZ background $\textrm{d} y/\textrm{d}z$ per logarithmic halo mass interval (blue scale), where each redshift is normalized separately. The black line shows the halo mass above and below which the halos contribute to $50\%$ of $\textrm{d} y/\textrm{d}z$ at a given redshift; it is slightly below the darkest blue scale as the distribution is skewed. The gray dashed line shows the nonlinear mass defined by $\sigma(M_*,z)=1.69$. The gray dash--dotted line shows the median mass of the main halos in the merger trees of $M_{\rm 500}(z=0) > 10^{14}$ $\rm M_{\odot}$ clusters from \cite{2017ApJ...844L..23C}. At each redshift, the mean tSZ background is dominated by a population of relatively rare and massive halos.}
    \label{fig:halo_mass}
\end{figure}

Strictly speaking, the direct observable in the clustering redshift technique is not $dy/dz$ but $\textrm{d}y/\textrm{d}z \times b_y$, where $b_{y}$ is the large-scale clustering bias of the test sample $b_{\rm T}$ in Equation~\ref{eq:wbar_to_dJdz}. In the halo model formalism, $b_{y}$ can be modeled as the ``Compton $y$-weighted halo bias'':
\begin{equation}
b_y(z) =
\frac{\int {\rm d}M \frac{{\rm d}n}{{\rm d}M} \tilde{y}_{0}(M,z)b_{\rm lin}(M,z)}
{\int {\rm d}M \frac{{\rm d}n}{{\rm d}M} \tilde{y}_{0}(M,z)},
\label{eq:by}
\end{equation}
where $b_{\rm lin}(M,z)$ is the linear halo bias given in \cite{tinker/etal:2010}. We note that $b_{y}$ is uniquely determined in the halo model, as it does not depend on the only free parameter $B$. In contrast, the tSZ amplitude $\textrm{d}y/\textrm{d}z$ depends on $B$ and has to be constrained observationally. The halo model $b_{y}$ correction is accurate to the $\sim10\%$ level due to the uncertainty in the theoretical halo mass function and halo bias.  In Appendix~\ref{appendix:b_y} we discuss $b_{y}$ in more detail and show the predicted values in Figure~\ref{fig:by}. 

\begin{figure*}[t!]
    \begin{center}
         \includegraphics[width=0.85\textwidth]{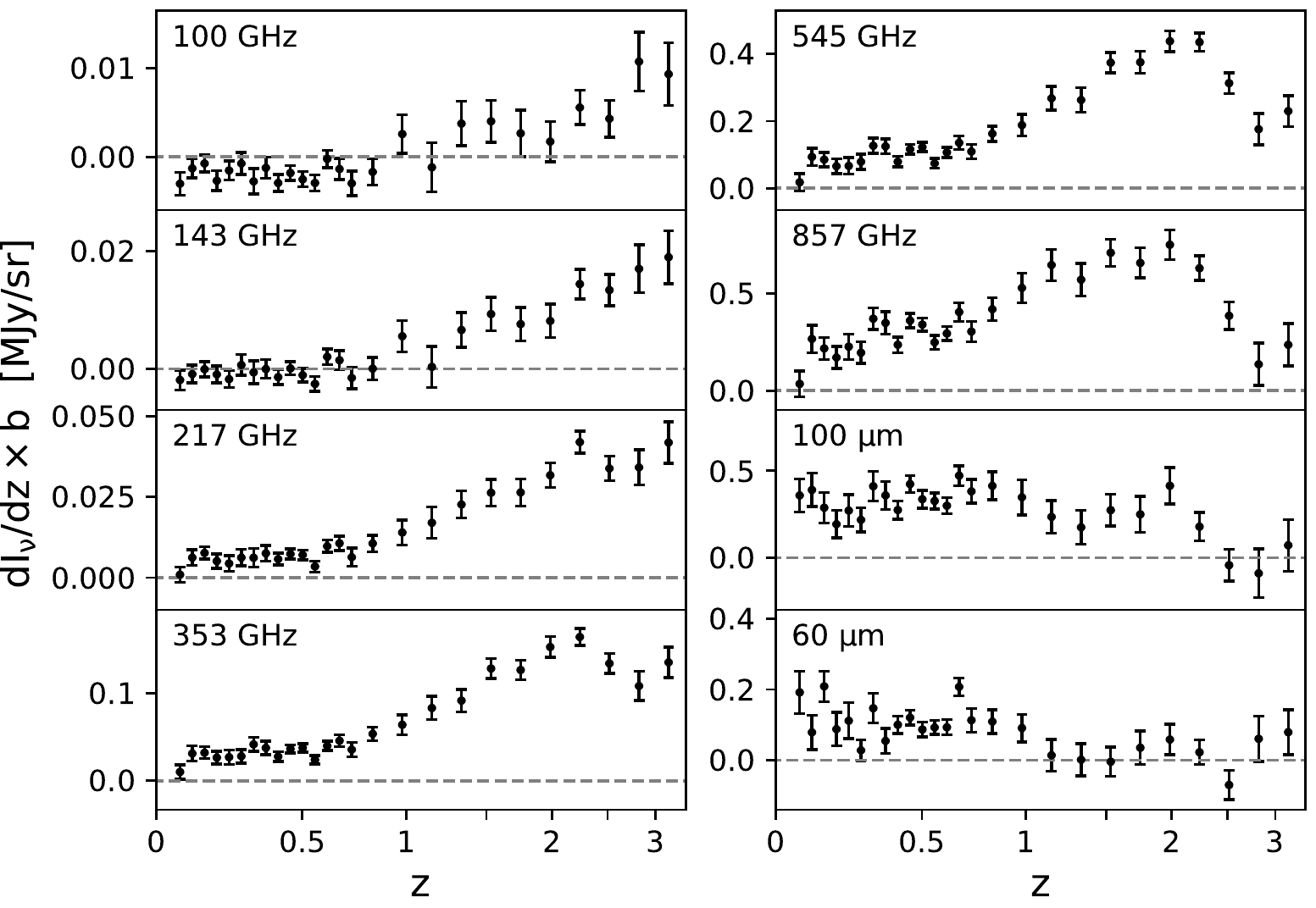}
    \end{center}
    \caption{Redshift-deprojected EBL intensities modulated by the clustering bias of the EBL for the eight {\it Planck} (100 to 857~GHz) and {\it IRAS} (100 and 60~$\mu m$) channels. These are measured by tomographically cross-correlating the {\it Planck} and {\it IRAS} intensity maps with two million spectroscopic redshifts in SDSS. Overall, the EBL redshift and spectral features are dominated by that of the CIB, while we clearly find decrements at $100$~GHz at $z<1$ originating from the cosmic tSZ effect background.}
    \label{fig:dIdz_b}
\end{figure*}

What kinds of structures dominate the cosmic mean tSZ background? Figure~\ref{fig:halo_mass} shows, in blue scale, the differential $\textrm{d} y/\textrm{d}z$ per dex of $M_{\rm 500}$ normalized to the total $\textrm{d} y/\textrm{d}z$ integrated down to $10^{8}\ h^{-1}$ $\textrm{M}_{\odot}$ at a given redshift. The black line shows the evolution of the $50\%$ $\textrm{d} y/\textrm{d}z$ mass, which corresponds to the median of the blue scale distribution at each redshift. Since the result is normalized, it depends only weakly on the mass bias $B$ (we use $B=1.3$ in the figure). We find that the ${\rm d}y/{\rm d}z$ is dominated by the most massive halos over about 2 dex in halo mass at a given epoch. This is a combined result from halo abundance and the steep mass scaling of the tSZ contribution from an individual halo: $\tilde{y}_0 \propto M^{5/3+\alpha_p}$. Cluster size halos with $M_{\rm 500}(z) > 10^{14}$ $\rm M_{\odot}$ contribute to $\sim60\%$ of $\textrm{d} y/\textrm{d}z$ at $z=0$, while galaxy groups with $M_{\rm 500}(z) = 10$$\mathrm{^{13-14}}$ $\rm M_{\odot}$ take over at $z\sim1$ as the cluster abundance decreases rapidly toward high redshifts. The exact same plot also applies to physical quantities that scale with $\textrm{d} y/\textrm{d}z$, which includes $\langle{P_{\rm{e}}}\rangle$, $\overbar{T}_{\rm{e}}$, and $\Omega_{\rm th}$. For comparison, we also show in Figure~\ref{fig:halo_mass} the characteristic ``nonlinear mass'' $\rm M_*$ and the mass of the core halos in protoclusters from \cite{2017ApJ...844L..23C}, which is obtained via taking the median of the main halo mass in the merger trees of a sample of $M_{\rm 500}(z=0) > 10^{14}$ $\rm M_{\odot}$ clusters. 

\begin{figure}[b!]
    \begin{center}
         \includegraphics[width=0.47\textwidth]{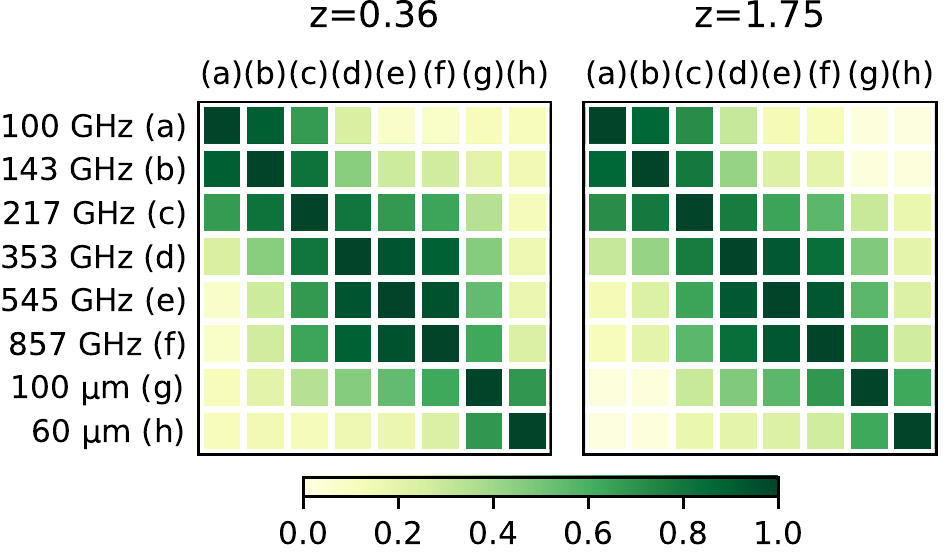}
    \end{center}
    \caption{Frequency correlation coefficient matrices for $\textrm{d}I_{\nu}/\textrm{d}z\times b$ at two selected redshift bins. Those at other redshifts are quantitatively  similar.}
    \label{fig:covariance}
\end{figure}

\begin{figure*}[t!]
    \begin{center}
         \includegraphics[width=0.942\textwidth]{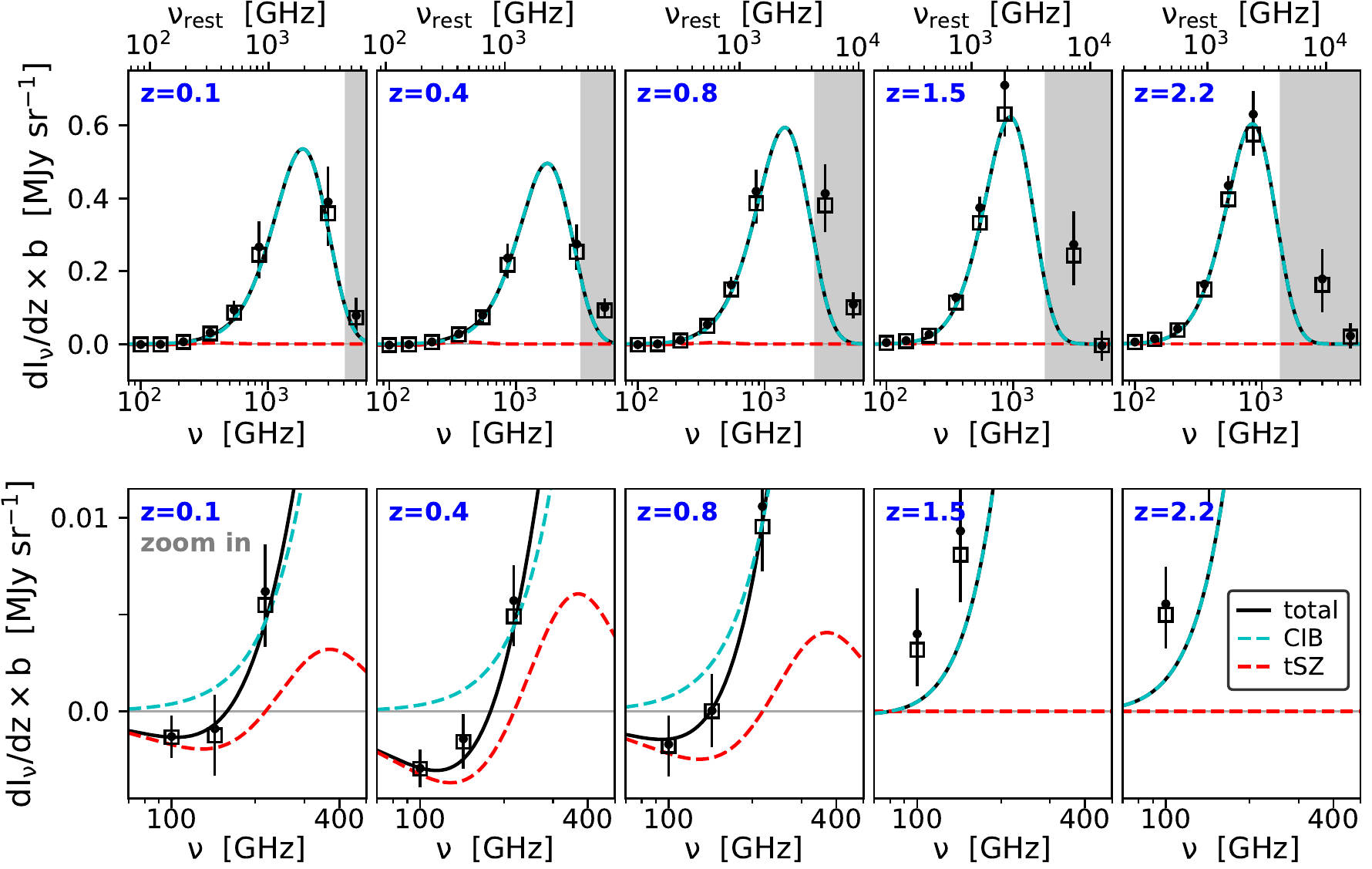}
    \end{center}
    \caption{Redshift-deprojected EBL SEDs and the two-component model fitting at five selected redshift bins. The lower panels show the zoom-ins to the tSZ frequency range. Filled data points show the intensities measured assuming $\nu\,I_\nu = \rm constant$ within the passbands, while open squares show those after color corrections. Black solid, cyan, and red dashed lines show the best-fitting total, CIB, and tSZ signals, respectively. Gray areas indicate the rest-frame frequencies above 4600~GHz ($\lambda < 65~\mu\rm m$), where our simple spectral model does not have enough freedom to describe hot dust in the CIB; data therein are thus not used in the SED fitting.}
    \label{fig:SED}
\end{figure*}

\section{Thermal SZ Background}
\label{sec:tSZ_background}

\subsection{Tomographic EBL Intensities}
\label{sec:results-dIdz}
We perform clustering-based redshift inference for the eight {\it Planck} and {\it IRAS} channel maps in 26 redshift bins up to $z\sim 3$. Figure~\ref{fig:dIdz_b} shows the results with the matter clustering $\overline{w}_{\rm DM}$ and reference bias $b_{\rm R}$ in Equation~\ref{eq:wbar_to_dJdz} already corrected. Here the $y$-axis, $\textrm{d}I_{\nu}/\textrm{d}z\times b$, is the observed EBL intensity decomposed into that emitted per unit redshift interval modulated by its clustering bias $b$ ($ b_{\rm T}$ in Equation~\ref{eq:wbar_to_dJdz}). 

Despite the complexity added by $b$, in Figure~\ref{fig:dIdz_b} we can identify some key features in the spectrum and energy budget of the EBL. First, at $z<0.8$, we see negative $\textrm{d}I_{\nu}/\textrm{d}z\times b$, that is, decrements, at 100~GHz. This is an unambiguous signature of the global tSZ effect as the CIB cannot be negative. In the same 100~GHz channel, $\textrm{d}I_{\nu}/\textrm{d}z\times b$ increases toward high redshifts, and the EBL appears to be taken over by the CIB in emission. At 857~GHz the redshift distribution shows a shape similar to that of the cosmic star-formation history \citep{Madau_Dickinson_2014}, and indeed the CIB is an excellent tracer of the dust-obscured star formation activities. We also find that, over the entire spectral range, the peak of $\textrm{d}I_{\nu}/\textrm{d}z\times b$ is shifted to higher redshifts at lower frequencies. This is because the CIB receives a more negative $K$-correction in lower frequency bands \citep[e.g.,][]{2005ARA&A..43..727L}. 

The bias factor $b$ in the $y$-axis quantifies the effective clustering properties of the EBL with respect to matter on large scales; it includes both the tSZ and CIB contributions, while the former is important only at low frequencies and low redshifts. The dominant CIB contribution of the bias $b$ is of order unity at low redshifts and increases by a factor of a few to $z\sim3$ \citep{2018A&A...614A..39M}. In the next subsection we will apply a simple per-redshift, two-component SED fitting to isolate the tSZ contribution.

As mentioned in Section~\ref{sec:clustering-redshift}, in a given band, the $\textrm{d}I_{\nu}/\textrm{d}z\times b$ measurements at different redshift bins are nearly independent. There is, however, a strong frequency covariance at a given redshift. This is visualized in Figure~\ref{fig:covariance}, which shows the correlation coefficient matrices at two redshift bins. We find that the structures of the frequency covariances are quantitatively similar throughout the entire redshift range considered (with the amplitudes scaled with the number of reference sources). This behavior is expected, as the noise is dominated by the Galactic foreground and the redshift-integrated CIB, which is largely stationary over different frequency bands.

\subsection{Snapshot SED Decomposition}
\label{sec:results-SED}
The power of the multichannel EBL tomography is the following: by slicing along the frequency channel axis at a fixed redshift in Figure~\ref{fig:dIdz_b}, we obtain a deprojected cosmic snapshot SED at that redshift bin. This redshift-dependent information would otherwise be lost in any map-level component separation analyses. For a physical interpretation, at each redshift we fit a two-component model consisting of the tSZ effect and the CIB thermal dust emission. This can be expressed as
\begin{eqnarray}
\frac{\textrm{d}I_{\nu}}{\textrm{d}z}\,b\,(\nu, z_i) &=& \left[ \frac{\textrm{d}y}{\textrm{d}z}\,b_y\,(z_i)\,G_y(\nu)\right]\nonumber\\
&+& \left[\frac{\textrm{d}I_{217}}{\textrm{d}z}(z_i)\,b_{\rm CIB}(\nu, z_i)\, G_{\rm CIB}(\nu, z_i)\right]\,,
\label{eq:tot_is_cib_plus_sz}
\end{eqnarray}
where each component consists of a redshift differential amplitude with no frequency dependence, a clustering bias $b$, and a spectral shape $G$. For both the tSZ effect and CIB, the redshift differential amplitude and bias are degenerate, so in our fitting we treat the product of the two as one effective amplitude parameter to be fitted at each redshift bin.

The spectral feature of the tSZ effect is unique. Assuming no relativistic correction, which is valid for all but the hottest clusters \citep{2018MNRAS.476.3360E}, there is no free parameter in the shape function $G_y(\nu) = g(x)\, I_{\nu,\,0}$ given in Equation~\ref{eq:delta_I_nu_given_y} and \ref{eq:g_x}. We thus have only one free parameter, $\textrm{d}y/\textrm{d}z \times b_y$, for the tSZ amplitude per redshift bin.

For the CIB, we adopt a single temperature modified blackbody spectrum in the optically thin regime \citep[e.g., ][]{2014A&A...571A..30P}:
\begin{equation}
G_{\rm CIB}(\nu) \propto \nu^{\beta}\,B_{\nu}(T, \nu)\,, 
\label{eq:MBB}
\end{equation}
where $\beta$ is the spectral index of the dust opacity or emissivity, and $B_{\nu}$ is the Planck function. With this model we have two shape parameters $T$ and $\beta$ plus one normalization parameter per redshift bin. For convenience, we choose to normalize the CIB amplitude at the observer-frame 217~GHz where the tSZ effect does not contribute. The clustering bias $b_\textrm{CIB}$ depends on redshift \citep{2018A&A...614A..39M} and also weakly on frequency \citep{2018MNRAS.475.3974W}. In this work we assume that the two dependencies are separable, that is, $b_\textrm{CIB} = b_\textrm{CIB}^z (z) \times b_\textrm{CIB}^{\nu}(\nu)$, where the first term is absorbed in our effective 217~GHz normalization and the second term, if significant, is absorbed into the best-fitting $\beta$ in the shape function $G_{\rm CIB}$. For these reasons, although we cannot empirically break the degeneracy between the CIB bias and other CIB parameters, it is not expected to affect the extraction of the tSZ amplitudes in the SED fitting.

We note that this simple modified blackbody spectrum drops exponentially at the Wien side, which is likely insufficient to model the potential hot dust emission in the mid-infrared \citep[e.g.,][]{2012MNRAS.425.3094C}. However, for the purpose of this work, it is sufficient to require the CIB parameterization to provide a reasonable baseline continuum at the Rayleigh Jeans tail for extraction of the tSZ amplitudes. We will explore the full constraining power of our CIB measurements using a more flexible spectral model in a forthcoming paper.

When combining the tSZ and CIB SED models, the tSZ is an effect in the observer frame, while the CIB is radiation emitted in the rest frame, which will appear redshifted. This means that the way CIB impacts the tSZ measurements is redshift-dependent: at $z=0$ a 20~K CIB peaks at $\sim2000$~GHz, while at $z=3$ it has shifted to the observed $\sim500$~GHz, which is more likely to affect the tSZ measurements.

\begin{figure*}[t!]
    \begin{center}
         \vspace{0.2cm}
         \includegraphics[width=0.7\textwidth]{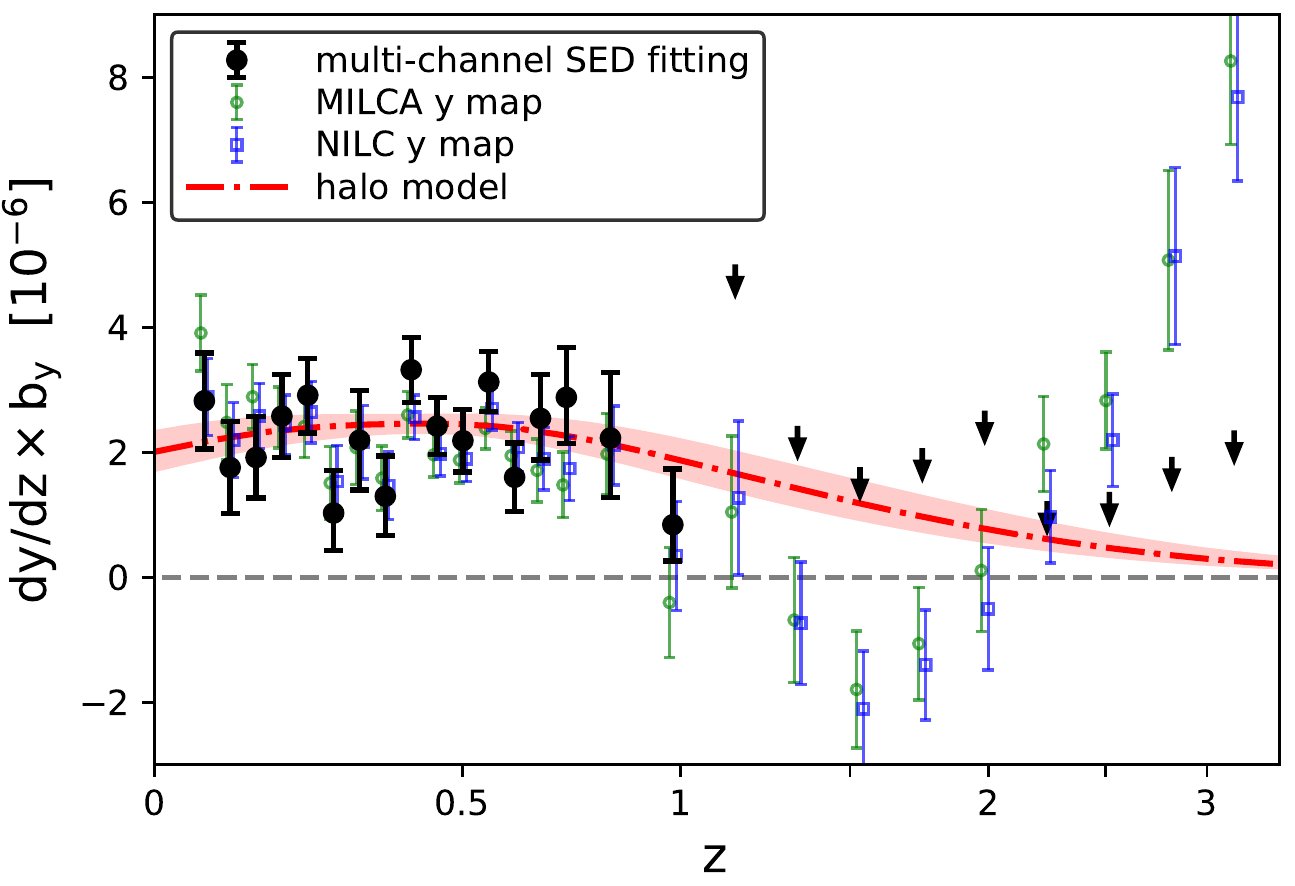}
    \end{center}
    \caption{Tomographic tSZ amplitudes $\textrm{d}y/\textrm{d}z$ modulated by the clustering bias $b_y$ as a function of redshift. Black data points and upper limits (1$\sigma$) show our fiducial measurements via multichannel redshift deprojection plus per-redshift cosmic SED fitting. Green/blue data points show our comparison measurements using $\it Planck$'s MILCA/NILC $y$ maps and for which CIB contamination is evident at high redshifts. The red dash--dotted line shows the best-fitting halo model with a redshift-evolving mass bias parameter, jointly constrained by all deprojected channel intensities with $\nu_{\rm rest}<4600$~GHz at $z<1$. The corresponding 68\% CL range is shown in the shaded band.}
    \label{fig:dYdz_b}
\end{figure*}

Having specified the tSZ plus CIB model, we fit it to the measured SEDs independently at each redshift bin. A Bayesian inference is implemented with the Markov Chain Monte Carlo (MCMC) approach using the {\sf emcee} code \citep{2013PASP..125..306F}. To avoid insufficient modeling in the mid-infrared, data points with rest-frame frequencies above 4600~GHz ($\lambda < 65~\mu\rm m$) are not used. At a given redshift, four parameters (one for tSZ and three for CIB) are constrained by six or seven data points (varying as the mid-infrared exclusion is applied in the rest frame). We require $\textrm{d}y/\textrm{d}z \times b_y > 0$ because a negative $y$ is unphysical, and so is the clustering bias $b_y$ under any reasonable halo models. For the CIB parameters, we set flat priors of $10~{\rm K}<T<40~{\rm K}$ and $0.5<\beta<3$. As our map intensities are quoted using the {\it IRAS} convention assuming an in-band spectrum of $\nu\,I_\nu = \rm constant$, during the SED fitting process we simultaneously fit the amplitudes of the color corrections following \cite{2014A&A...571A...9P}. 

Figure~\ref{fig:SED} shows the EBL spectra for five selected redshift bins, obtained from the $\textrm{d}I_{\nu}/\textrm{d}z\times b$ data points given in Figure~\ref{fig:dIdz_b}. The zoom-ins to the tSZ frequency range are shown on the lower panels. The two-component SED fitting results are overlaid. Filled data points and open squares denote the $\textrm{d}I_{\nu}/\textrm{d}z\times b$ measurements before and after the simultaneously fitted color corrections, respectively. The red and cyan dashed lines show the best-fitting tSZ and the CIB spectra, respectively, whose sum makes up the total SED shown in the black solid lines. We find that the CIB dominates the cosmic SED at frequencies above 400~GHz over the entire redshift range, with a broad thermal peak corresponding to a nearly constant temperature of $\sim 22$~K. At $z<0.8$, the tSZ effect contributes significantly, while the best-fitting $\textrm{d}y/\textrm{d}z \times b_y$ at higher redshifts tend to hit the zero bound. 

In the $z=1.5$ and $z=2.2$ panels (and at most other $z\gg1$ bins not shown here), there appears to be an excess above the best-fitting CIB line at the rest-frame 200--600~GHz. This results in higher reduced chi-square values and reveals that it is insufficient to describe the high-redshift CIB SED using the one-component modified blackbody CIB model. We find that this rest-frame submillimeter and millimeter CIB excess at high redshifts is similar to that seen in the Small Magellanic Cloud \citep{2012ApJ...757..103D}. This might suggest a more diverse dust population, optical properties, or energetics in both individual galaxies and the CIB than previously thought. 

Extractions of the tSZ amplitudes are unavoidably CIB-model-dependent at high redshifts. Fortunately, at $z < 0.8$ this is not the case: as the 100~GHz intensities are significantly negative and the 217~GHz intensities are close to zero, the SZ decrements that we detect would not be accounted for by any reasonable scenario of the CIB, which is strictly positive in intensity. The low-redshift tSZ measurements would thus be robust. We treat our $\textrm{d}y/\textrm{d}z \times b_y$ constraints differently in two regimes. At $z<1$ we report the measurements. At $z>1$ we acknowledge the model dependence and report the upper limits as our current CIB model underestimates the CIB intensities at frequencies where we expect tSZ decrements. 

\setlength{\tabcolsep}{8pt}
\renewcommand{\arraystretch}{1.12}
\begin{table*}[t]
\centering
\caption{\label{table:measurement_values}}
Sunyaev-Zeldovich Effect Background Amplitudes from multichannel EBL Tomography\\
\vspace{0.2cm}
\begin{tabular}{ccccccccc}
\hline
\hline
$z$ & ($\textrm{d}y/\textrm{d}z) b_y$ & $\textrm{d}y/\textrm{d}z$ & $\langle b\,P_{\rm e}\rangle$ & $\overbar{T}_{\rm{e}}$ & $k_{\rm{B}}\overbar{T}_{\rm{e}}$ & $\rho_{\rm{th}}$ & $\rm \Omega_{th}$ & $B$ \\

 & $[10^{-6}]$ & $[10^{-6}]$ & [meV cm$^{-3}$] & [$10^6$ K] & [keV] & [10$^{-5}$ eV cm$^{-3}$] & [10$^{-8}$] &  \\
 \hline
0.07 & $2.82\,^{+0.76}_{-0.77}$ & $0.94\,^{+0.26}_{-0.26}$ & $0.17\,^{+0.05}_{-0.05}$ & $2.51\,^{+0.68}_{-0.68}$ & $0.22\,^{+0.06}_{-0.06}$ & $9.23\,^{+2.5}_{-2.51}$ & $1.93\,^{+0.52}_{-0.53}$ & $1.14\,^{+0.22}_{-0.14}$ \\
0.11 & $1.75\,^{+0.73}_{-0.74}$ & $0.58\,^{+0.24}_{-0.24}$ & $0.11\,^{+0.05}_{-0.05}$ & $1.47\,^{+0.62}_{-0.62}$ & $0.13\,^{+0.05}_{-0.05}$ & $5.41\,^{+2.27}_{-2.27}$ & $1.13\,^{+0.47}_{-0.48}$ & $1.5\,^{+0.58}_{-0.26}$ \\
0.14 & $1.91\,^{+0.65}_{-0.65}$ & $0.63\,^{+0.21}_{-0.21}$ & $0.13\,^{+0.04}_{-0.04}$ & $1.52\,^{+0.52}_{-0.51}$ & $0.13\,^{+0.04}_{-0.04}$ & $5.58\,^{+1.9}_{-1.88}$ & $1.17\,^{+0.4}_{-0.39}$ & $1.44\,^{+0.37}_{-0.22}$ \\
0.18 & $2.58\,^{+0.67}_{-0.66}$ & $0.84\,^{+0.22}_{-0.21}$ & $0.19\,^{+0.05}_{-0.05}$ & $1.93\,^{+0.5}_{-0.5}$ & $0.17\,^{+0.04}_{-0.04}$ & $7.1\,^{+1.84}_{-1.82}$ & $1.49\,^{+0.39}_{-0.38}$ & $1.23\,^{+0.22}_{-0.15}$ \\
0.22 & $2.91\,^{+0.59}_{-0.61}$ & $0.94\,^{+0.19}_{-0.19}$ & $0.22\,^{+0.05}_{-0.05}$ & $2.07\,^{+0.42}_{-0.43}$ & $0.18\,^{+0.04}_{-0.04}$ & $7.6\,^{+1.53}_{-1.58}$ & $1.59\,^{+0.32}_{-0.33}$ & $1.16\,^{+0.16}_{-0.12}$ \\
0.27 & $1.03\,^{+0.68}_{-0.6}$ & $0.33\,^{+0.22}_{-0.19}$ & $0.08\,^{+0.06}_{-0.05}$ & $0.69\,^{+0.46}_{-0.4}$ & $0.06\,^{+0.04}_{-0.03}$ & $2.54\,^{+1.69}_{-1.48}$ & $0.53\,^{+0.35}_{-0.31}$ & $2.04\,^{+1.28}_{-0.49}$ \\
0.31 & $2.19\,^{+0.79}_{-0.8}$ & $0.69\,^{+0.25}_{-0.25}$ & $0.19\,^{+0.07}_{-0.07}$ & $1.4\,^{+0.51}_{-0.51}$ & $0.12\,^{+0.04}_{-0.04}$ & $5.15\,^{+1.86}_{-1.87}$ & $1.08\,^{+0.39}_{-0.39}$ & $1.36\,^{+0.39}_{-0.22}$ \\
0.36 & $1.29\,^{+0.65}_{-0.62}$ & $0.41\,^{+0.2}_{-0.2}$ & $0.12\,^{+0.06}_{-0.06}$ & $0.79\,^{+0.39}_{-0.38}$ & $0.07\,^{+0.03}_{-0.03}$ & $2.89\,^{+1.44}_{-1.39}$ & $0.6\,^{+0.3}_{-0.29}$ & $1.81\,^{+0.76}_{-0.35}$ \\
0.4 & $3.32\,^{+0.51}_{-0.52}$ & $1.03\,^{+0.16}_{-0.16}$ & $0.32\,^{+0.05}_{-0.05}$ & $1.92\,^{+0.3}_{-0.3}$ & $0.17\,^{+0.03}_{-0.03}$ & $7.04\,^{+1.09}_{-1.11}$ & $1.47\,^{+0.23}_{-0.23}$ & $1.07\,^{+0.11}_{-0.08}$ \\
0.45 & $2.42\,^{+0.46}_{-0.46}$ & $0.74\,^{+0.14}_{-0.14}$ & $0.25\,^{+0.05}_{-0.05}$ & $1.33\,^{+0.25}_{-0.25}$ & $0.11\,^{+0.02}_{-0.02}$ & $4.88\,^{+0.93}_{-0.92}$ & $1.02\,^{+0.19}_{-0.19}$ & $1.28\,^{+0.16}_{-0.12}$ \\
0.5 & $2.18\,^{+0.5}_{-0.5}$ & $0.66\,^{+0.15}_{-0.15}$ & $0.24\,^{+0.06}_{-0.06}$ & $1.15\,^{+0.26}_{-0.26}$ & $0.1\,^{+0.02}_{-0.02}$ & $4.21\,^{+0.97}_{-0.96}$ & $0.88\,^{+0.2}_{-0.2}$ & $1.34\,^{+0.22}_{-0.15}$ \\
0.55 & $3.12\,^{+0.48}_{-0.47}$ & $0.94\,^{+0.15}_{-0.14}$ & $0.37\,^{+0.06}_{-0.06}$ & $1.56\,^{+0.24}_{-0.24}$ & $0.13\,^{+0.02}_{-0.02}$ & $5.74\,^{+0.89}_{-0.87}$ & $1.2\,^{+0.19}_{-0.18}$ & $1.08\,^{+0.11}_{-0.08}$ \\
0.61 & $1.6\,^{+0.55}_{-0.54}$ & $0.48\,^{+0.16}_{-0.16}$ & $0.2\,^{+0.07}_{-0.07}$ & $0.77\,^{+0.26}_{-0.26}$ & $0.07\,^{+0.02}_{-0.02}$ & $2.81\,^{+0.96}_{-0.96}$ & $0.59\,^{+0.2}_{-0.2}$ & $1.56\,^{+0.44}_{-0.24}$ \\
0.66 & $2.54\,^{+0.71}_{-0.67}$ & $0.75\,^{+0.21}_{-0.2}$ & $0.35\,^{+0.1}_{-0.09}$ & $1.16\,^{+0.32}_{-0.31}$ & $0.1\,^{+0.03}_{-0.03}$ & $4.28\,^{+1.19}_{-1.12}$ & $0.89\,^{+0.25}_{-0.23}$ & $1.18\,^{+0.22}_{-0.15}$ \\
0.72 & $2.88\,^{+0.79}_{-0.74}$ & $0.84\,^{+0.23}_{-0.22}$ & $0.42\,^{+0.12}_{-0.11}$ & $1.26\,^{+0.35}_{-0.33}$ & $0.11\,^{+0.03}_{-0.03}$ & $4.64\,^{+1.28}_{-1.2}$ & $0.97\,^{+0.27}_{-0.25}$ & $1.08\,^{+0.19}_{-0.14}$ \\
0.82 & $2.23\,^{+1.05}_{-0.96}$ & $0.64\,^{+0.3}_{-0.27}$ & $0.37\,^{+0.17}_{-0.16}$ & $0.91\,^{+0.43}_{-0.39}$ & $0.08\,^{+0.04}_{-0.03}$ & $3.35\,^{+1.58}_{-1.44}$ & $0.7\,^{+0.33}_{-0.3}$ & $1.19\,^{+0.44}_{-0.23}$ \\
0.98 & $0.84\,^{+0.89}_{-0.58}$ & $0.24\,^{+0.25}_{-0.16}$ & $0.16\,^{+0.17}_{-0.11}$ & $0.31\,^{+0.33}_{-0.22}$ & $0.03\,^{+0.03}_{-0.02}$ & $1.15\,^{+1.22}_{-0.8}$ & $0.24\,^{+0.25}_{-0.17}$ & $1.84\,^{+1.48}_{-0.6}$ \\
1.15 & $<4.55$ & $<1.25$ & $<1.07$ & $<1.55$ & $<0.13$ & $<5.69$ & $<1.19$ & $>0.67$ \\
1.33 & $<1.97$ & $<0.52$ & $<0.56$ & $<0.61$ & $<0.05$ & $<2.25$ & $<0.47$ & $>0.96$ \\
1.53 & $<1.31$ & $<0.34$ & $<0.45$ & $<0.37$ & $<0.03$ & $<1.38$ & $<0.29$ & $>1.06$ \\
1.75 & $<1.61$ & $<0.41$ & $<0.67$ & $<0.42$ & $<0.04$ & $<1.56$ & $<0.33$ & $>0.85$ \\
1.98 & $<2.21$ & $<0.54$ & $<1.11$ & $<0.54$ & $<0.05$ & $<1.97$ & $<0.41$ & $>0.62$ \\
2.24 & $<0.77$ & $<0.18$ & $<0.47$ & $<0.17$ & $<0.01$ & $<0.62$ & $<0.13$ & $>0.98$ \\
2.52 & $<0.91$ & $<0.21$ & $<0.68$ & $<0.18$ & $<0.02$ & $<0.68$ & $<0.14$ & $>0.76$ \\
2.82 & $<1.47$ & $<0.32$ & $<1.34$ & $<0.27$ & $<0.02$ & $<1.0$ & $<0.21$ & $>0.49$ \\
3.15 & $<1.89$ & $<0.38$ & $<2.11$ & $<0.32$ & $<0.03$ & $<1.16$ & $<0.24$ & $>0.35$ \\
\hline
\end{tabular}
\end{table*}

\subsection{Tomographic tSZ Amplitudes}
\label{sec:tSZ-history}

\subsubsection{\texorpdfstring{$\textrm{d}y/\textrm{d}z \times b_y$}{dy/dz x by}}

Figure~\ref{fig:dYdz_b} shows our measurements of $\textrm{d}y/\textrm{d}z \times b_y$, the bias-weighted mean tSZ amplitude, based on our multichannel EBL tomography and the snapshot SED fitting. We find secure detections at $z<1$ (black data points) and place upper limits (1$\sigma$; black arrows) at $z>1$ where the CIB contamination becomes strong. We summarize all of the values in Table~\ref{table:measurement_values}\footnote{Electronic version available at \url{https://github.com/yikuanchiang/tSZ-tomography}.}. We find $\textrm{d}y/\textrm{d}z \times b_y$ of order $10^{-6}$. The redshift evolution of $\textrm{d}y/\textrm{d}z \times b_y$ is mild and is consistent with that of the best-fitting halo model shown in the red dash--dotted line and the shaded band. 

For comparison, we also perform redshift tomography on \textit{Planck}'s MILCA and NILC $y$ maps in exactly the same manner as that for individual channel maps. The results are shown in Figure~\ref{fig:dYdz_b} with the green and blue open data points, respectively, with small offsets in $x$-axis added by hand for clarity. At $z<1$, the $y$ map results are consistent with our fiducial multichannel results, which suggests that the \textit{Planck} $y$ maps are reasonably free of CIB contamination in this regime. 
The high degree of covariance seen for the black, blue, and green data points at $z<1$ is expected as these measurements are largely based on the same intensity map data from \textit{Planck}. At high redshifts, we find that the $y$ map results deviate significantly from the multichannel-based results. The $\textrm{d}y/\textrm{d}z \times b_y$ values from {\it Planck} $y$ maps are unphysically negative at $z\sim 1.5$ and increase and overshoot dramatically at $z\sim3$. These anomalies indicate that the CIB contamination in the $y$ maps is significant at high redshifts.

How exactly do the NILC and MILCA algorithms pick up unwanted CIB signals? During the construction of the $y$ maps, both algorithms use a set of scale- and frequency-dependent weights to perform a linear combination of the channel intensity maps. The weights are determined by requiring a unit response to $y$ and zero response to CMB, and minimizing the variance of the reconstructed $y$ field. As a result, the CIB enters not only by adding noise but also by biasing the $y$ field as it correlates with $y$ both spatially and spectrally in a redshift-dependent manner. A possible reason for the unphysical, negative $\textrm{d}y/\textrm{d}z \times b_y$ seen at $z\sim1.5$ is that the $y$ maps are forced to compensate for the overestimation of the tSZ signals at $z\sim3$ such that the integrated $y$ is unbiased. Due to the CIB contamination in the $\it Planck$ $y$ maps, hereafter we only discuss our fiducial result based on the multichannel approach.

\begin{figure}[b!]
    \begin{center}
         \includegraphics[width=0.462\textwidth]{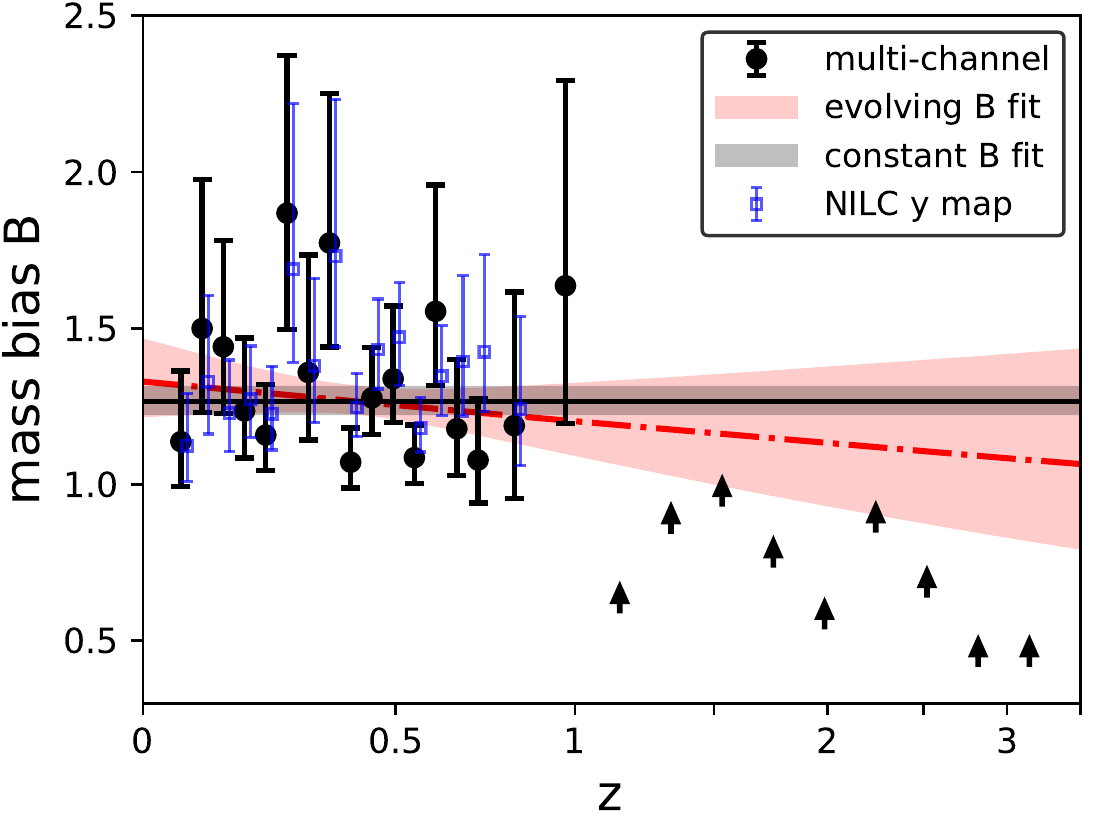}
    \end{center}
    \caption{Best-fitting mass bias parameter, $B$, constrained by our tomographic tSZ measurements. Symbols and legend are the same as those shown in Figure~\ref{fig:dYdz_b}. The redshift-evolving and constant-$B$ models are shown in the red and black bands, respectively.}
    \label{fig:mass_bias}
\end{figure}

\subsubsection{Constraining the Mass Bias}
\label{sec:B-fitting}
We compare our multichannel $\textrm{d}y/\textrm{d}z \times b_y$ measurements with those predicted in the halo model (Section~\ref{sec:halo-model}) and constrain free parameter(s) therein. Recall that in our halo model, the (mass-dependent) pressure profile, halo mass function, and linear halo bias are all fixed. The observable $\textrm{d}y/\textrm{d}z \times b_y$ at a given redshift thus only depends on the unknown mass bias parameter $B$, with a simple scaling of $\textrm{d}y/\textrm{d}z\propto B^{-5/3-\alpha_p}$, where $\alpha_p = 0.12$. A higher observed $\textrm{d}y/\textrm{d}z \times b_y$ thus implies a lower $B$, which then corresponds to a lower degree of non-thermal pressure support in halos. For each $\textrm{d}y/\textrm{d}z \times b_y$ data point in Figure~\ref{fig:dYdz_b}, we calculate the corresponding $B$ using the halo model and show the result in Figure~\ref{fig:mass_bias} (also summarized in Table~\ref{table:measurement_values}). For the multichannel approach (black data points and limits), each per-redshift posterior of $B$ is directly converted from the MCMC sampling of the corresponding $\textrm{d}y/\textrm{d}z \times b_y$ posterior. For the direct $\it{Planck}$ $y$ map tomography, the per-redshift mean and 1-sigma estimates for $\textrm{d}y/\textrm{d}z \times b_y$ (Figure~\ref{fig:dYdz_b}) are converted into those for $B$ in Figure~\ref{fig:mass_bias}, showing only the $\rm NILC$ map results for clarity. The errors in $B$ are noticeably non-Gaussian due to its nonlinear relationship with the observable $\textrm{d}y/\textrm{d}z \times b_y$.
We find that $B$ does not evolve with redshift significantly at least up to $z\sim1$, while it is less clear at $z>1$, where only one-sided limits are obtained. 

\begin{figure*}[t!]
    \begin{center}
         \includegraphics[height=0.385\textwidth]{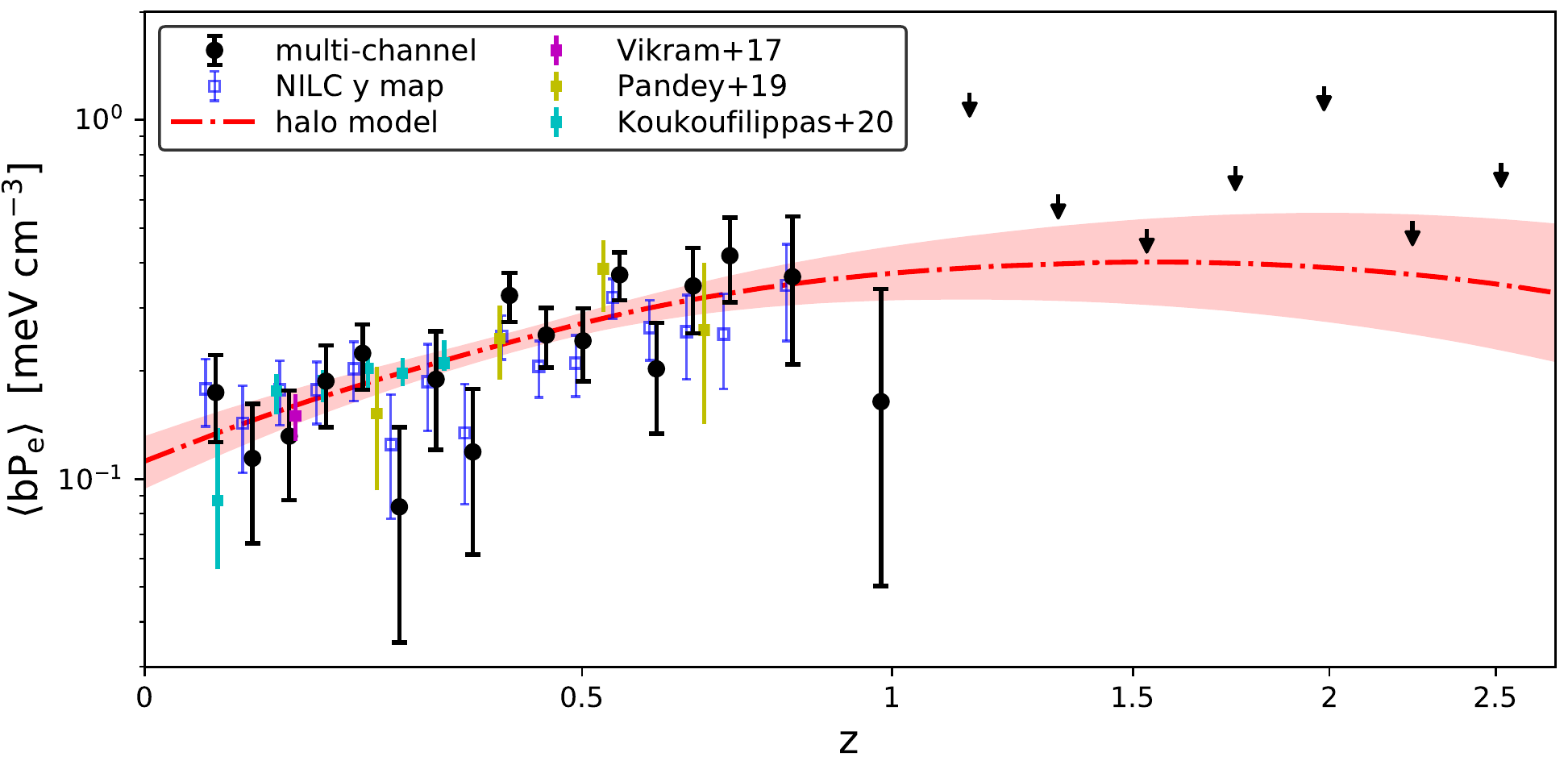}
    \end{center}
    \caption{Halo bias-weighted mean electron pressure of gas in the universe as a function of redshift. Black and blue data points/limits and the red bands show our measurements and halo-model fit corresponding to those of the same symbols in Figure~\ref{fig:dYdz_b}. Magenta, yellow, and cyan data points show previous measurements presented in the literature \citep{2017MNRAS.467.2315V,2019PhRvD.100f3519P,2020MNRAS.491.5464K}. All of the data points, limits, and model fit are in agreement within the uncertainties.}
    \label{fig:bPe}
\end{figure*}

The halo model allows us to extrapolate our mean tSZ constraints at $z<1$ to higher redshifts, where the strong CIB makes robust detections of the tSZ challenging. To properly propagate the uncertainty in $B$, we perform a joint spectral-redshift fitting in addition to the per-redshift SED fitting presented in Section~\ref{sec:results-SED}. In the joint fitting, all $\textrm{d}I_{\nu}/\textrm{d}z\times b$ measurements in Figure~\ref{fig:dIdz_b} at $z<1$ with the rest-frame frequency below $4600$ GHz ($\lambda_{\rm rest}>65$ $\mu$m) are used simultaneously. The same tSZ plus CIB spectral models introduced in Section~\ref{sec:results-SED} are used, and we assume a smooth power-law evolution with $(1+z)$ for all model parameters up to $z=1$. These include the tSZ mass bias $B$, CIB normalization $b_{\rm CIB}\,I_{217}$, and CIB shape parameters $T$ and $\beta$. The same MCMC code is used to sample the posteriors assuming uninformative priors for all of the parameters. Over the redshift range of $0<z<1$, we find a mass bias of $B(z) = (1.33\pm0.13) \times (1+z)^{0.15\pm 0.25}$. The other posteriors for the CIB parameters and the full covariance matrix are shown in Appendix~\ref{appendix:joint_fitting_posterior}. Consistent with the per-redshift fit, the mass bias jointly constrained at $z<1$ shows no significant redshift evolution, which is plotted in Figure~\ref{fig:mass_bias} as the red dash--dotted line and shaded band over the entire redshift range. The corresponding $\textrm{d}y/\textrm{d}z \times b_y$ is shown in the same way in Figure~\ref{fig:dYdz_b}. For comparison, we also repeat the $z<1$ joint fitting assuming a constant mass bias, finding $B=1.27^{+0.05}_{-0.04}$, which is shown as the black line and shaded band in Figure~\ref{fig:mass_bias}. The mass biases constrained at $z<1$ with and without redshift evolution are consistent with each other, and are both are ompatible with the per-redshift 1-$\sigma$ lower limits at $z>1$. 

\subsubsection{\texorpdfstring{$\langle{bP_{\rm{e}}}\rangle$}{<bPe>}}

We now compare our measurements with those presented in the literature. The tSZ tomographic quantity $\textrm{d}y/\textrm{d}z \times b_y$ directly constrains $\langle{bP_{\rm{e}}}\rangle$, the halo bias-weighted mean electron pressure in the universe (Equation~\ref{eq:bPe}). This is the quantity usually reported in the literature, which we show in Figure~\ref{fig:bPe}. As labeled in the figure, we show the results obtained by \cite{2017MNRAS.467.2315V}, a single-epoch stacking of the $\it Planck$ $y$ maps at the locations of galaxy groups at $z\sim 0.15$, and tomographic tSZ measurements from cross-correlating the $\it Planck$ $y$ maps with photometric-redshift galaxies in \cite{2019PhRvD.100f3519P} and \cite{2020MNRAS.491.5464K}.
The results of our analysis are shown with black data points and limits. We point out that, in addition to covering a much wider redshift range, our analysis explicitly handles redshift-dependent CIB contamination and, in contrast to previous studies using photometric redshifts, our data points are quasi-independent thanks to the spectroscopic redshift references we used. The numerical values of our results are presented in Table~\ref{table:measurement_values}. The overall agreement between all of the measurements, using different datasets and estimates of the tSZ signals, supports the robustness of the observational basis of the cosmic tSZ history. We will provide a physical interpretation of these results in section~\ref{sec:discussion_Omega_th}.

\subsubsection{Interpreting the Mass Bias}
\label{sec:mass_bias_discuss}
Our results, supported by the agreement with previous measurements presented in the literature, lead to a best-fitting mass bias parameter $B(z) = (1.33\pm0.13) \times (1+z)^{0.15\pm 0.25}$ and $B=1.27^{+0.05}_{-0.04}$ in scenarios with and without redshift dependence in $B$, respectively (Figure~\ref{fig:mass_bias}). This corresponds to another commonly used parameter $1-b=B^{-1}=0.79\pm0.03$ for the nonevolving case. Thus, cluster masses determined by combining \cite{arnaud/etal:2010} and \cite{planck_inter_v:2013} using resolved X-ray and tSZ observations assuming hydrostatic equilibrium are, on average, about $20\%$ lower than the true masses.

If non-physical calibration issues are negligible, the excess of $B$ from unity can be attributed to non-thermal pressure support in halos. The magnitude of $B$ we found is consistent with that in cosmological hydrodynamical simulations and analytic predictions of structure formation where halos are additionally supported by internal bulk motions and turbulence sourced by hierarchical mass assembly
\citep{dolag/etal:2005,iapichino/niemeyer:2008,vazza/etal:2006,vazza/etal:2009,vazza/etal:2016,vazza/etal:2018,lau/kravtsov/nagai:2009,maier/etal:2009,shaw/etal:2010,iapichino/etal:2011,2012ApJ...758...74B,2014ApJ...782..107N,2014MNRAS.442..521S,2015MNRAS.448.1020S,2016MNRAS.455.2936S,angelinelli/etal:2020}. This implies that additional kinetic energy injections from baryonic feedback, mostly from active galactic nucleus activities, are either small (because, e.g., they are confined in the small volume of galaxy cluster cores) or largely thermalized. Our mass bias parameter is roughly consistent with those found in the Compton $y$-galaxy cross-correlation literature (\citealt{2019PhRvD.100f3519P, 2020MNRAS.491.5464K, 2020PASJ...72...26M}), but also note that \citealt{2018MNRAS.480.3928M} reported a higher value using low-redshift 2MASS galaxies. The mass bias parameters obtained via combining the primary CMB and the tSZ autopower spectrum or cluster counts are somewhat higher at $B\sim 1.6-1.7$
\citep{planck_sz_cosmo:2014,2016A&A...594A..24P,hurier/lacasa:2017,2018MNRAS.477.4957B,2020MNRAS.497.1332B,osato/etal:2018,osato/etal:2020,salvati/etal:2018,salvati/etal:2019}. The discrepancy in the reported mass bias is not highly significant at face value but could be appreciable, considering that the measurements are not entirely independent as they are based on part of the same multifrequency dataset. The $\it Planck$ $y$ maps were used in all aforementioned measurements, while we show in Figure~\ref{fig:dYdz_b} that these maps are strongly affected by the CIB at $z>1$. This could introduce systematics especially for studies relying heavily on the projected autopower spectrum of Compton $y$ if the impact of the CIB is not fully taken into account. 

Finally, we point out that, if in the future $B$ can be precisely estimated, observational constraints on $\langle{bP_{\rm{e}}}\rangle$ can then become useful in constraining cosmological parameters; in particular, the amplitude scales as $\sigma_8 (\Omega_m/B)^{0.4}\, h_0^{-0.21}$, as shown by \cite{2018MNRAS.477.4957B}.

\begin{figure*}[t!]
    \begin{flushright}
         \includegraphics[height=0.5\textwidth]{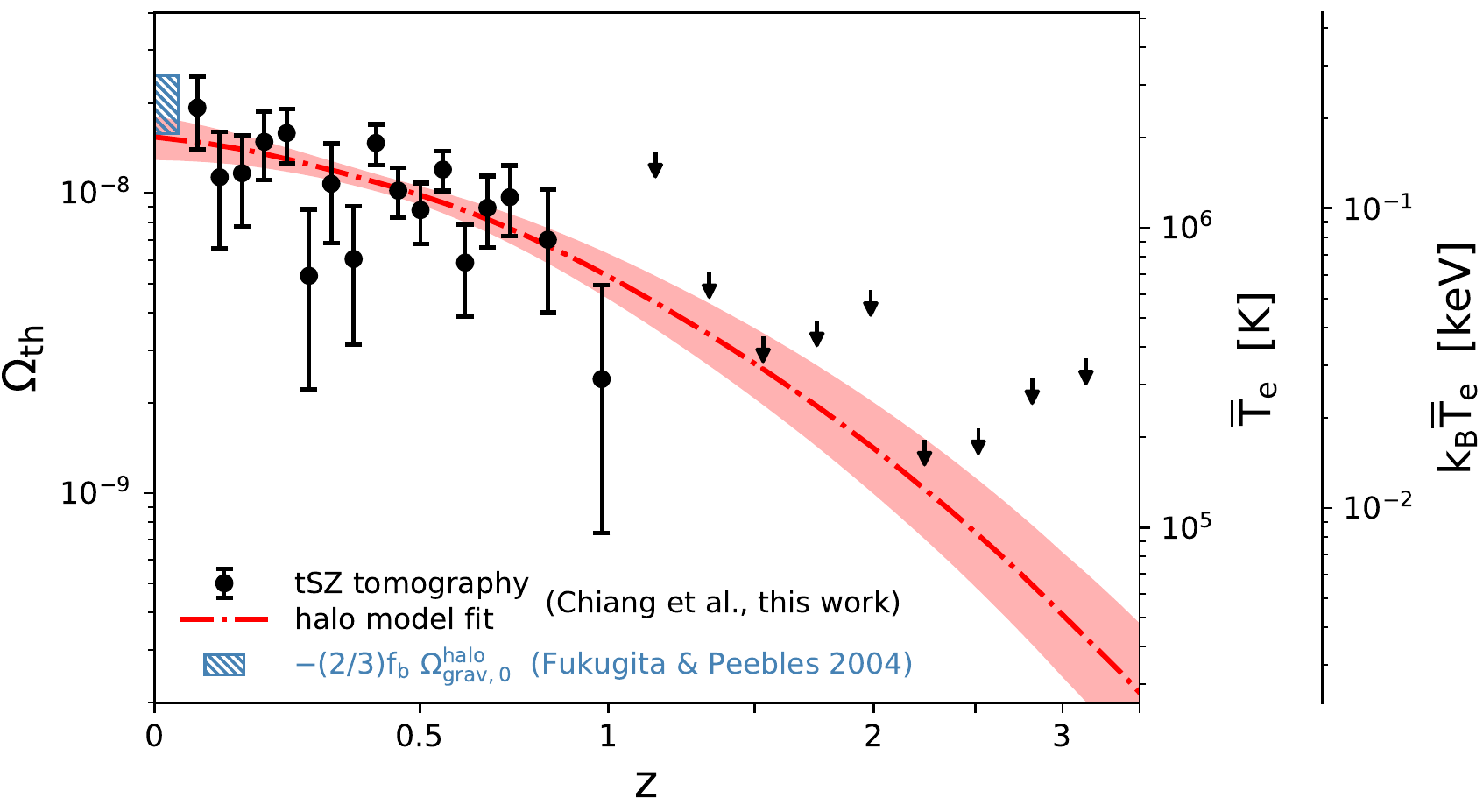}
    \end{flushright}
    \caption{Cosmic thermal energy density parameter $\Omega_{\rm th} = \rho_{\rm th}/\rho_{\rm crit}$ as a function of redshift. This is equivalent to the density-weighted mean gas temperature $\overbar{T}_{\rm{e}}$ \citep{1999ApJ...514....1C,2000PhRvD..61l3001R} labeled in the right-hand $y$-axes. Black data points and upper limit (1$\sigma$) are from our multichannel-based tomographic tSZ measurements. The red dash--dotted line and shaded band show the best-fitting halo model and the 68\% CL range jointly constrained by the tomographic channel intensities at $z<1$. If the halo gas is entirely virialized and thermalized, we would expect $\Omega_{\rm th} = \Omega'_{\rm grav}$, where $\Omega'_{\rm grav} = - (2/3) f_{\rm b}\ \Omega_{\rm grav}^{\rm halo}$ with $f_{\rm b} = 0.157$ being the cosmic baryon mass fraction and $\Omega_{\rm grav}^{\rm halo}$ is the gravitational binding energy in halos. We plot the present-day $\Omega'_{\rm grav}$ in the steel blue hatched region using $\Omega_{\rm grav}^{\rm halo}$ calculated in \cite{2004ApJ...616..643F}. If $\Omega_{\rm th}$ is significantly lower than $\Omega'_{\rm grav}$, the deficit could be attributed to non-thermal energy in halos.}
    \label{fig:omega_th}
\end{figure*}

\subsubsection{tSZ Sky Monopole}
\label{sec:total_y}
The precision measurement of the sky-averaged, redshift-integrated monopole Compton $y$ parameter offers a powerful integral constraint on the thermal history of the universe. The Far Infrared Absolute Spectrophotometer (FIRAS) on the Cosmic Background Explorer gave the upper bound on the monopole $y<1.5\times 10^{-5}$ \citep[95\%~CL;][]{1996ApJ...473..576F}.
Theoretical estimates based on halo models and cosmological hydrodynamic simulations are in the range of (1--2$)\times 10^{-6}$ \citep{1996A&A...314...13B,2000MNRAS.317...37D, 2000PhRvD..61l3001R, 2001MNRAS.327L...5N,2001PhRvD..63f3001S,2004MNRAS.355..451Z,2010ApJ...725...91B,2015PhRvL.115z1301H,dolag/komatsu/sunyaev:2016}; thus, the expected level of the mean $y$ is only one order of magnitude smaller than the FIRAS upper bound. 

It is possible to improve upon the FIRAS limit by several orders of magnitude with future spectral distortion experiments based on technology already available \citep{2011JCAP...07..025K,2014JCAP...02..006A,2019arXiv190901593C}. While an absolutely calibrated spectrometer is needed to directly measure the monopole Compton $y$, one can use the anisotropies of $y$ to constrain its correlated component with the large-scale structure. \citet{2015JCAP...08..013K} reported a limit of monopole $y<2.2\times 10^{-6}$ (95\%~CL) from the probability density function of {\it Planck}'s Compton $y$ maps, but the unconstrained map zero-points (because {\it Planck} is not an absolutely calibrated experiment) make the interpretation ambiguous. In our analysis, we can avoid this ambiguity by measuring the total $y$ correlated with spectroscopic reference tracers of the cosmic web.

Using the results presented in Section~\ref{sec:tSZ-history}, we can evaluate the redshift-integrated $y$ directly detected in our measurements up to $z=1$. Correcting for $b_y$ using the halo model (Section~\ref{sec:halo-model}) and integrating our data points over redshift by taking $\Sigma_{z=0}^{1}\, \textrm{d}y/\textrm{d}z|_{z=z_i}\, \Delta z_i$, we find $y(0<z<1) = (6.7\pm 0.6) \times 10^{-7}$.
The only assumption made for this measurement is the $y$-weighted halo bias $b_y$, which can be calculated robustly (Section~\ref{sec:halo-model}). It gives a robust lower bound for the redshift-integrated mean Compton $y$ parameter that would be measured by future missions.

By integrating and extrapolating our best-fitting halo-model $\textrm{d}y/\textrm{d}z$ beyond the redshifts directly probed, we find a total cosmic $y = 1.22^{+0.23}_{-0.17}\times10^{-6}$ and $y = 1.12 \pm 0.07\times10^{-6}$ for the evolving and constant mass bias scenarios, respectively; the two are consistent within the uncertainty, and a maximum redshift of 6 is sufficient for both integrals to converge. This estimate of the total cosmic $y$ provides an important baseline for future spectral distortion missions. Comparing the halo-model-extrapolated total $y$ with the direct sum at $z<1$, we conclude that we have directly detected about $60\%$ of the monopole Compton $y$ of the cosmic tSZ background from large-scale structure.

\section{Cosmic Thermal Energy Density}
\label{sec:discussion_Omega_th}

We now present our empirical constraints for the comoving thermal energy density in the universe $\rho_{\rm th}$ (Equation~\ref{eq:rho_th}) and the corresponding energy density parameter $\Omega_{\rm th}$ (Equation~\ref{eq:omega_th}). These are shown in Figure~\ref{fig:omega_th} using our measurements in black as well as the best-fit halo model and its 68\% CL range (allowing redshift-evolving mass bias) in red.
Our detections and upper limits allow us to probe the growth of cosmic thermal energy density over more than an order of magnitude, during a period of about 12 Gyr. This is an outcome of cosmic structure formation: matter density fluctuations seeded in the early universe are amplified by gravity and form collapsed halos; baryons then follow and get shock heated to the virial temperatures. About $70\%$ of the growth of the present-day $\rho_{\rm th}$ occurred over 8 Gyr of cosmic time between $z=1$ and $z=0$, where we have direct detections. The growth rate of $\rho_{\rm th}$, however, decreases below $z\sim1$, which can be more clearly seen in the halo-model fit. This is consistent with the picture where the growth of structures is slowed down by the late-time accelerated expansion of the universe because of dark energy. Note that the halo-model fit uses only our tSZ detections at $z<1$. Nonetheless, the upper limits we obtained at $1<z<3$ are fully compatible with while pushing the upper bound of the extrapolation of the halo-model fit. The limits thus meaningfully support the change in structure growth rate that is due to the transition from matter to dark energy domination.
At the present time, we find
\begin{equation}
\Omega_{\rm th} = (1.5\pm0.3) \times 10^{-8}~~{\rm at~}z=0\;.
\end{equation}
This reservoir of energy is due to the conversion of gravitational potential energy (dominated by dark matter) into heat carried by baryons. As shown in Figure~\ref{fig:halo_mass}, the dominant contribution of $\Omega_{\rm th}$ originates from  galaxy clusters and protoclusters. For such structures, cooling (mainly through X-rays) is restricted to the central region of clusters and takes places on a timescale that is longer than the Hubble time. The thermal energy $\Omega_{\rm th}$ therefore represents an energy reservoir that accumulates with time without suffering from significant losses. It is a robust probe of structure formation.

Over the years, the thermal history of the universe has been described in different manners. 
\cite{1999ApJ...514....1C} and \cite{2000PhRvD..61l3001R} attempted to characterize it using hydrodynamic simulations through the quantity $\overbar{T}_{\rm{e}}$, the cosmic mean density-weighted gas temperature, which we show using the right axes in Figure~\ref{fig:omega_th}. An early attempt to constrain this quantity at $z=0$ has been made using the WMAP data \citep{2004PhRvD..69h3524A}. We note that $\overbar{T}_{\rm{e}}$ provides an effective temperature of baryons in the universe, but it should not be interpreted as an effective virial temperature to derive a characteristic halo mass, as a significant fraction of cosmic baryons are outside halos.

To our knowledge, the thermal energy density was first introduced in \cite{2004MNRAS.355..451Z}, who provided predictions based on simulations, which are in good agreement with our measurements up to $z\sim3$. The cosmic energy inventory compiled by \cite{2004ApJ...616..643F} does not specifically include the thermal energy, but these authors describe the gravitational binding energy in halos, $\Omega_{\rm grav}^{\rm halo}$, which is a related quantity because the thermal history of the universe is almost entirely driven by gravitational collapse of cosmic structures. It is therefore interesting to compare $\Omega_{\rm th}$ with $\Omega_{\rm grav}^{\rm halo}$. The present-day value of the $\Omega_{\rm grav}^{\rm halo}$ is calculated in \cite{2004ApJ...616..643F} for two halo mass bins corresponding to galaxy clusters and $\rm L_{*}$ galaxies, where the former contributes a factor of a few more. If halos are fully virialized and baryons within them are entirely in the ionized gas phase and fully thermalized, we would expect $\Omega_{\rm th} = \Omega'_{\rm grav}$, where $\Omega'_{\rm grav} = - (2/3) f_{\rm b}\ \Omega_{\rm grav}^{\rm halo}$ with $f_{\rm b} = 0.157$ being the cosmic baryon mass fraction. Additional non-thermal pressure supports would lead to $\Omega_{\rm th} < \Omega'_{\rm grav}$. We plot $\Omega'_{\rm grav}$ at $z=0$ in Figure~\ref{fig:omega_th} using the steel blue hatched region by summing up the cluster and $\rm L_{*}$ host contributions from \cite{2004ApJ...616..643F}. The height of the hatched region shows a $\pm25\%$ range, which roughly corresponds to the uncertainty of the dark matter statistics used in \cite{2004ApJ...616..643F}. We see that at the present day, $\Omega_{\rm th}$ matches their 2004 estimate of $\Omega'_{\rm grav}$ remarkably well. 

In a companion paper (\citealt{chiang/etal:prep-2}), we will present an update of the \cite{2004ApJ...616..643F} calculation of $\Omega_{\rm grav}^{\rm halo}$ using the improved understanding and characterization of dark matter statistics. A more detailed comparison between the cosmic gravitational and thermal  energy budgets will be presented as a function of redshift, and we will estimate the non-thermal contribution.

\section{Conclusion}
\label{sec:conclusion}
By measuring redshift-dependent amplitudes of the mean thermal Sunyaev-Zeldovich (SZ) effect background, we have obtained new constraints on the thermal history of the universe. The cosmic thermal energy content is dominated by hot gas in galaxy clusters at low redshifts and groups and protoclusters at high redshifts. Its evolution is almost entirely driven by the growth of structures as baryons get shock heated in collapsing dark matter halos.

To probe this thermal history, we employ the clustering-based redshift inference technique to extract cosmic time-dependent SZ signals. Our analysis is based on a set of angular cross-correlations between eight sky intensity maps in the $\it{Planck}$ and Infrared Astronomical Satellite missions with two million spectroscopic redshift references in the Sloan Digital Sky Surveys. It consists of the following steps:
\begin{itemize}
    \item We first derive a set of snapshot SEDs for the far-infrared to microwave background light as a function of redshift up to $z\sim3$.
    \item We decompose these snapshot SEDs into the SZ and thermal dust components using well-defined spectral models. 
    \item We obtained direct observational constraints on $\langle bP_{\rm e} \rangle$, the halo bias-weighted mean electron pressure of the universe up to $z\sim 3$, with detections up to $z\sim1$, the highest redshift reached to date. 
\end{itemize}
We have used these $\langle b P_{\rm e}\rangle$ estimates to derive the mean thermal pressure
$\langle P_{\rm e}\rangle=\langle b P_{\rm e}\rangle/b_y$, where $b_y$ is the SZ-weighted halo bias computed using the halo model. This allows us to probe the following:

\begin{itemize}
    \item We estimate $\overbar{T}_{\rm{e}}$, the density-weighted electron temperature of the universe, which rises from from $7\times 10^5~{\rm K}$ at $z=1$ to $2\times 10^6~{\rm K}$ today.
    
    \item We probe the cosmic thermal history quantified by the evolution of $\Omega_{\rm th}$, the cosmic thermal energy density parameter. We find that $\Omega_{\rm th}$ grows by more than an order of magnitude since $z=3$ and reaches $1.5 \times10^{-8}$ at the present time.
    \item We find the mass bias parameter of $\it{Planck}$'s universal pressure profile of $B=1.27$ (or $1-b=1/B=0.79$), consistent with the magnitude of non-thermal pressure in gas motion and turbulence from mass assembly. 
    \item We determine the redshift-integrated total Compton $y$ parameter of $1.22\times10^{-6}$, which will be tested by future spectral distortion experiments. About 60\% of this originates in the large-scale structure at $z<1$, which we detect directly.
\end{itemize}

In a companion paper (\citealt{chiang/etal:prep-2}), we will present a comparison of the cosmic thermal ($\Omega_{\rm th}$) and gravitational ($\Omega_{\rm grav}^{\rm halo}$) energy contents as a function of redshift. By combining these two energy budgets, we will infer the contribution originating from non-thermal processes.

\acknowledgments
This work was supported in part by NSF grant AST1313302 and NASA grant NNX16AF64G (Y.C., B.M.),
the Excellence Cluster ORIGINS, which is funded by the Deutsche Forschungsgemeinschaft (DFG, German Research Foundation) under Germany's Excellence Strategy - EXC-2094 - 390783311 (E.K.), and JSPS KAKENHI grant Nos. JP15H05896 (R.M., E.K.) and JP20K14515 (R.M.). The Kavli IPMU is supported by World Premier International Research Center Initiative (WPI), MEXT, Japan. 

\bibliography{references}

\appendix

\section{One-halo term impact in clustering redshifts}
\label{appendix:1_halo_impact}
\begin{figure*}[h!]
    \begin{center}
         \includegraphics[width=0.95\textwidth]{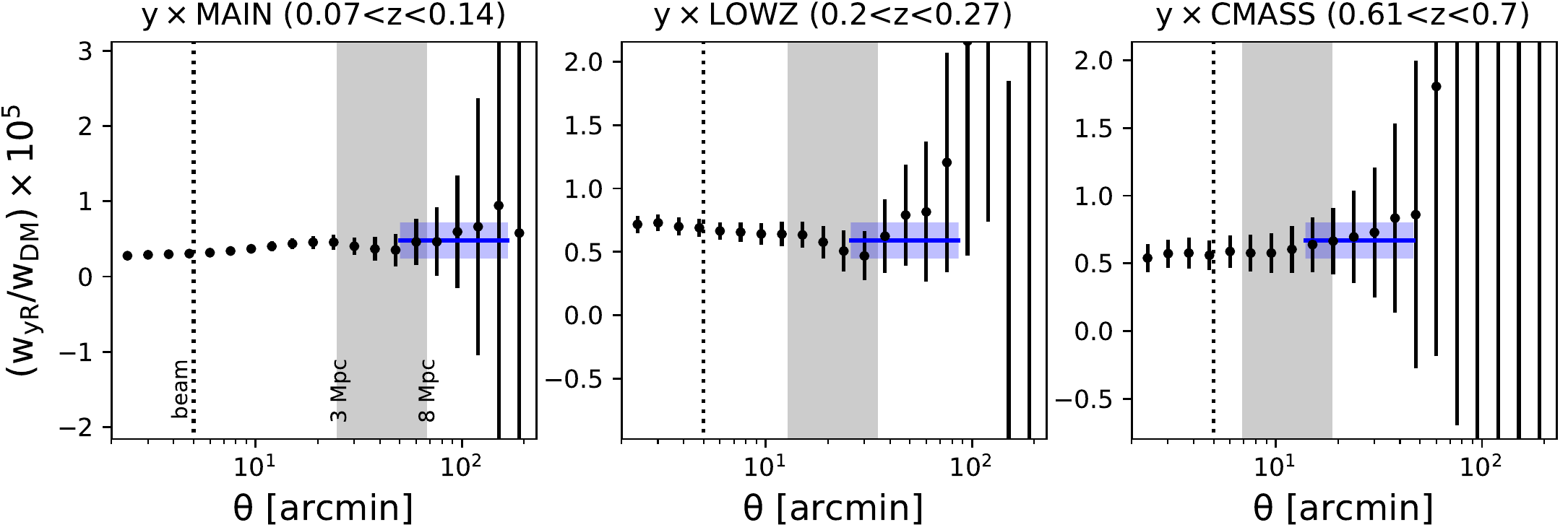}
    \end{center}
    \vspace{-0.295cm}
    \caption{Ratio between the $y$-reference cross-correlation function and the dark matter autocorrelation function at three redshift bins with the MAIN, LOWZ, and CMASS galaxy samples as the reference, respectively. The NILC $y$ map is used, with a half-width at half-maximum of the $5'$ beam indicated in the dotted lines. At each redshift bin, the blue band shows the best-fitting value to the 6--20 Mpc (physical) amplitudes. The gray shaded regions indicate the 3--8 Mpc regime used for our $\textrm{d}y/\textrm{d}z$ measurements, where the correlation function ratios do not show significant deviations from the 6--20 Mpc values.}
    \label{fig:1halo_term_impact}
\end{figure*}

The clustering redshift technique assumes that the cross-correlation function of the test and reference samples follows the linear relationship given in Equation~\ref{eq:wbar_to_dJdz}. With the Compton $y$ as the test sample, $\textrm{T} = y$, the estimation of $\textrm{d}y/\textrm{d}z$ could be biased if $\overbar{w}_{y\textrm{R}}$, the effective $y$-reference clustering amplitude, is measured at small scales where the one-halo term clustering dominates. Here we show that the potential one-halo term impact is not significant in our measurements. Figure~\ref{fig:1halo_term_impact} shows $w_{y\textrm{R}}/w_{\textrm{DM}}$ as a function of angular scale at three redshift bins using the MAIN, LOWZ, and CMASS galaxy samples as the reference, where $w_{y\textrm{R}}$ is the full $y$-reference angular correlation function and $w_{\textrm{DM}}$ is the theoretical dark matter autocorrelation function \citep[Equation~10 in][]{Chiang_2019}. To get higher signal-to-noise ratios, the {\it Planck} NILC $y$ map is used here instead of the fiducial multichannel approach. In the calculation of $w_{\textrm{DM}}$, we include the effect of the beam of the $y$ map (dotted lines) and a 10 deg high-pass filtering on large scale used also in the $w_{y\textrm{R}}$ measurements to suppress wide-angle systematics. The $w_{y\textrm{R}}/w_{\textrm{DM}}$ ratio is expected to be proportional to $\textrm{d}y/\textrm{d}z$ in the linear regime ($\gtrsim 10$ Mpc); to guide the eyes, we show a constant fit to the 6--20 Mpc (physical) amplitudes in the blue bands. The main $\textrm{d}y/\textrm{d}z$ result in this paper is based on clustering signals extracted over 3--8 Mpc (physical), which is indicated with the gray shaded regions in Figure~\ref{fig:1halo_term_impact}. We find that in all three redshift bins with different galaxies as the reference, the $w_{y\textrm{R}}/w_{\textrm{DM}}$ ratios at 3--8 Mpc do not show significant departures from the large-scale values. This suggests that given the precision we are reaching, our $\textrm{d}y/\textrm{d}z$ measurements are robust against the systematic because of the breakdown of Equation~\ref{eq:wbar_to_dJdz} on small scales, which is due to the scale-dependent bias of the Compton $y$ field or the reference sources.

\section{Thermal SZ-weighted halo bias \texorpdfstring{$\MakeLowercase{b_y}$}{by}}
\label{appendix:b_y}
\begin{figure}[h!]
    \begin{center}
         \includegraphics[width=0.465\textwidth]{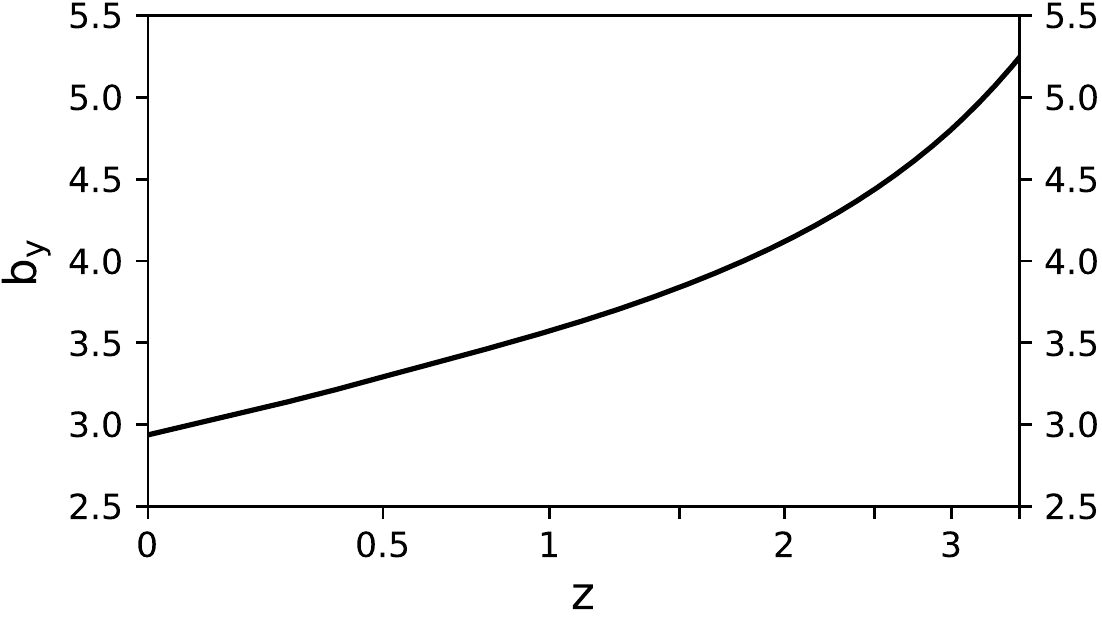}
    \end{center}
    \vspace{-0.33cm}
    \caption{Compton $y$-weighted halo bias $b_y$ in the halo model used to correct for the clustering-based observable $\textrm{d}y/\textrm{d}z \times b_y$ to get $\textrm{d}y/\textrm{d}z$.}
    \label{fig:by}
\end{figure}

In our tomographic tSZ measurements, the direct clustering-based observable is $\textrm{d}y/\textrm{d}z \times b_y$, where $b_y$ is the $y$-weighted halo bias. We thus need to calculate $b_y$ to obtain $\textrm{d}y/\textrm{d}z$. As it turns out, $b_y$ can be robustly predicted in the halo model (Section~\ref{sec:halo-model}) as it does not depend on the mass bias parameter $B$. To see this, we substitute $\tilde{y}_{0}$ in Equation~\ref{eq:by} using Equation~\ref{eq:pe}, \ref{eq:px}, and \ref{eq:mass_bias_B}, which gives
\begin{equation}
b_y(z) =
\frac{\int {\rm d}M \frac{{\rm d}n}{{\rm d}M} 
M^{5/3+\alpha_p} b_{\rm lin}(M,z)}
{\int {\rm d}M \frac{{\rm d}n}{{\rm d}M} M^{5/3+\alpha_p}}\,,
\label{eq:by_simple}
\end{equation}
with $M = M_{\rm 500}$. Although $B$ enters in both the numerator and denominator, it is canceled in the ratio. This expression also shows that $b_{y}$ can be understood as the $M^{5/3+\alpha_p}$-weighted halo bias (where $\alpha_p = 0.12$). Figure~\ref{fig:by} shows the predicted $b_{y}$ as a function of redshift, which is similar to the linear bias of massive halos that dominate the tSZ signals (Figure~\ref{fig:halo_mass}).

\section{Posteriors in joint SED fitting}
\label{appendix:joint_fitting_posterior}
To combine our tSZ constraining power across a range of redshift, in section~\ref{sec:B-fitting} we jointly fit all $\textrm{d}I_{\nu}/\textrm{d}z\times b$ measurements at $z<1$ and the rest-frame frequencies less than 4600~GHz. To obtain a realistic shape of the redshift evolution of $\textrm{d}y/\textrm{d}z$, we use the halo model (Section~\ref{sec:halo-model}) with an unknown mass bias parameter $B$ and fit a constant $B$ or a power-law $B(z) \propto (1+z)^{\gamma}$. This delivers the halo-model fit shown as the smooth lines and shaded bands in Figure~\ref{fig:dYdz_b}, \ref{fig:mass_bias}, \ref{fig:bPe}, \ref{fig:omega_th}. 
In these figures, the red labels show the fit with the power-law $B(z)$, while the constant $B$ fit is shown only in Figure~\ref{fig:mass_bias} because the results of the two are consistent within the uncertainty (so no significant redshift evolution of $B$ is found). 

\begin{figure*}[h!]
    \begin{center}
         \includegraphics[width=0.48\textwidth]{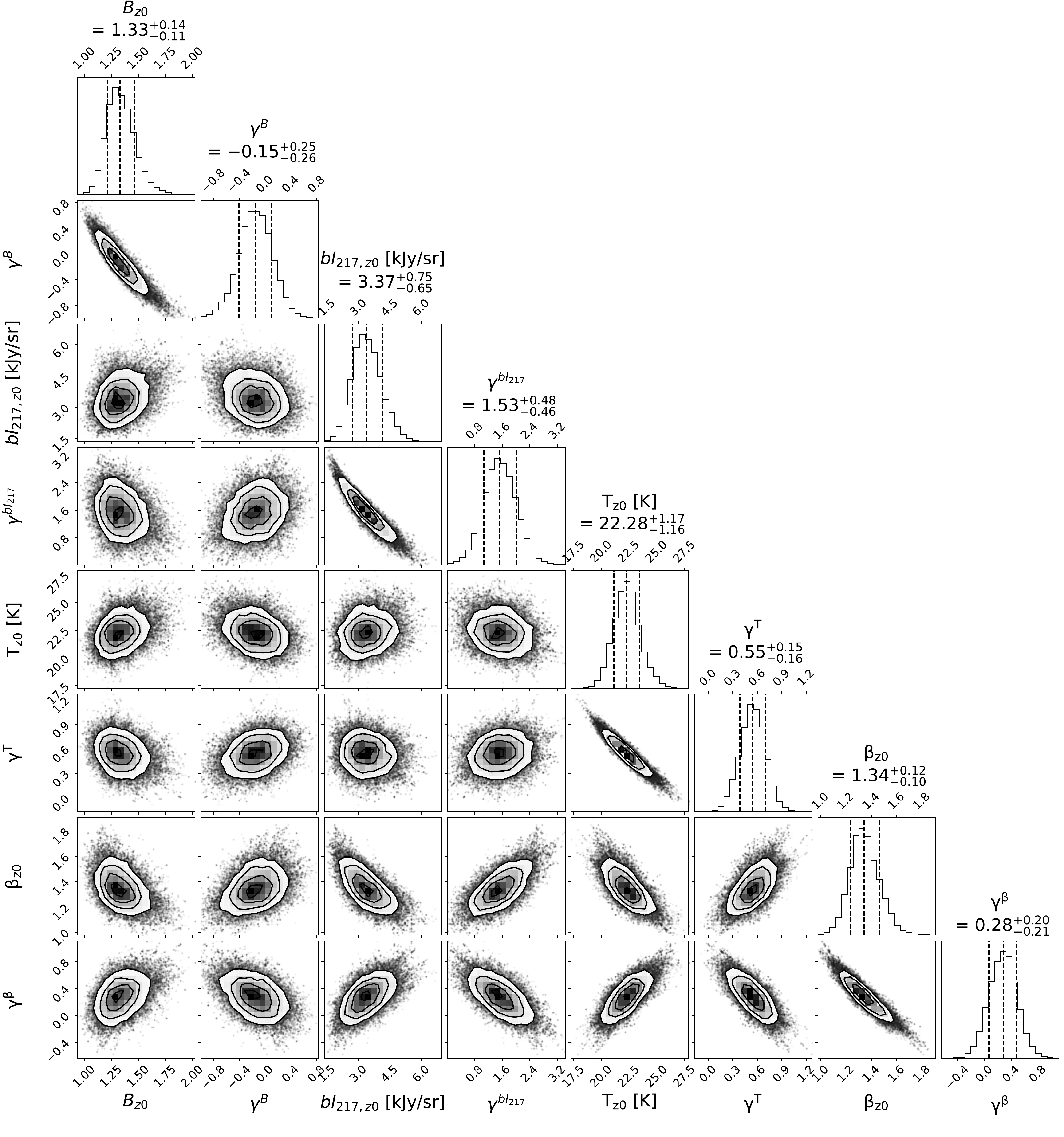}
         \hspace{0.25cm}
         \includegraphics[width=0.48\textwidth]{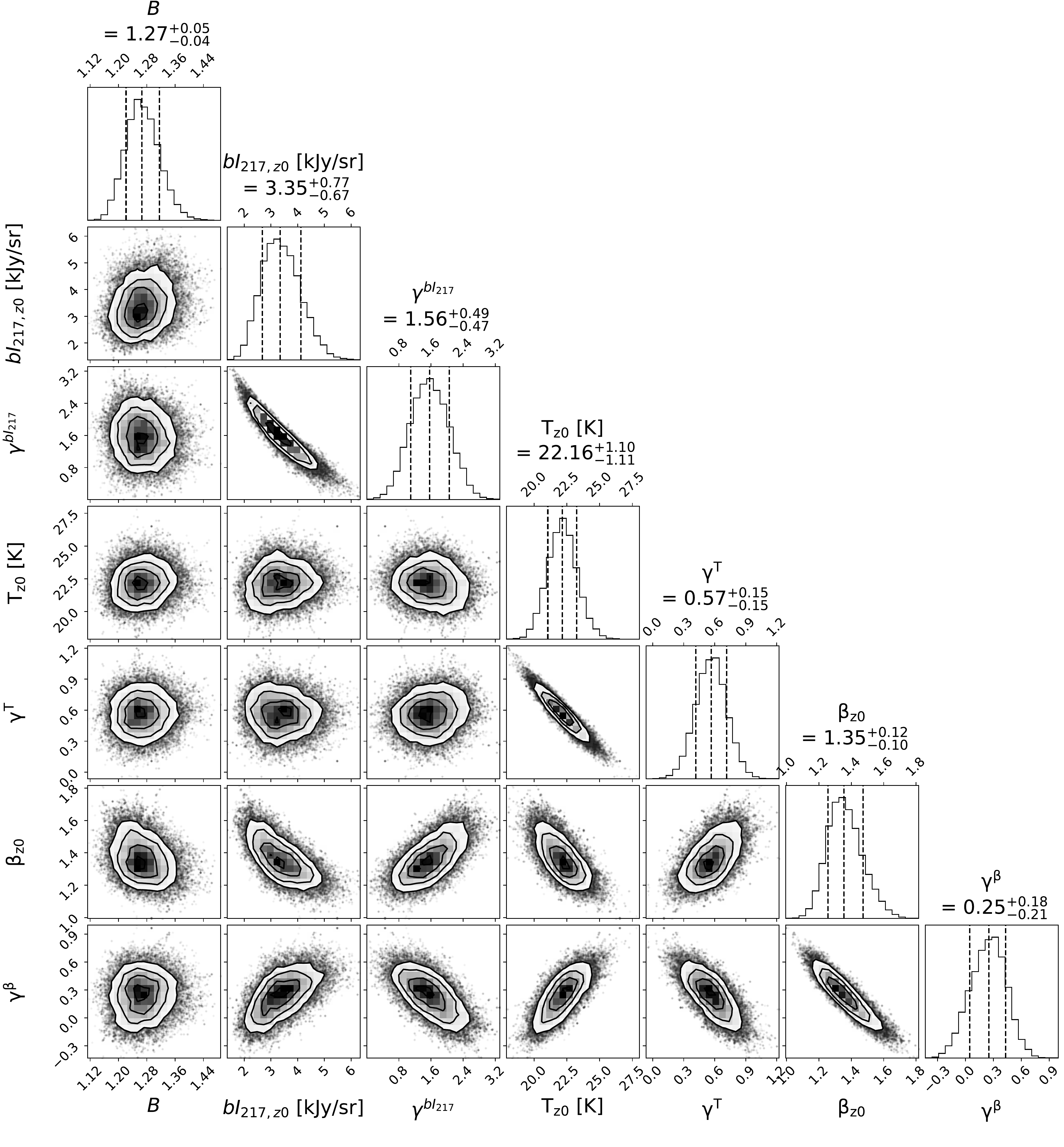}
    \end{center}
    \caption{Posteriors for the tSZ plus CIB  parameters fit to $\textrm{d}I_{\nu}/\textrm{d}z\times b$ measurements at $z<1$ and $\nu_{\rm rest}<4600$ GHz. The left panel shows that allowing evolving $B$ while the right panel assumes a constant $B$. Posteriors in the two scenarios are consistent within the uncertainties.}
    \label{fig:smooth_z_fit_trangle_plots}
\end{figure*}

The posteriors of $B$ are obtained by marginalizing over the CIB parameters, which are assumed to evolve as power laws of $1+z$ up to $z=1$. For the case allowing for an evolving $B$, we find $B(z) = (1.33\pm0.13) \times (1+z)^{0.15\pm 0.25}$, the CIB dust temperature of $T(z) = (22.3\pm1.2) \times (1+z)^{0.55\pm 0.15}$ K, the CIB dust opacity power index of $\beta(z) = (1.34\pm0.11) \times (1+z)^{0.28\pm 0.21}$, and the observer-frame 217~GHz normalization of the CIB of $b_{\rm CIB}\,I_{217}(z) = (3.37\pm0.70) \times (1+z)^{1.63\pm 0.47}$ $\rm kJy\ sr^{-1}$. The full posteriors and covariances for the evolving and nonevolving $B$ cases are shown in the left and right panels of Figure~\ref{fig:smooth_z_fit_trangle_plots}, respectively. The mass bias parameter $B$ is not degenerate with any of the CIB parameters, which supports the robustness of our constraints.

\end{document}